\documentclass[preprint,12pt]{elsarticle}
\usepackage[version=4]{mhchem}
\usepackage[T1]{fontenc}
\usepackage[utf8]{inputenc}
\usepackage{geometry}
\usepackage{verbatim}
\usepackage{amsmath}
\usepackage{amssymb}
\usepackage{graphicx}
\usepackage{setspace}
\usepackage{physics}
\usepackage{mathtools}
\usepackage[dvipsnames]{xcolor}
\usepackage[normalem]{ulem}
\usepackage{cancel}
\makeatletter
\providecommand{\tabularnewline}{\\}
\date{}

\usepackage[english]{babel} 

\@ifundefined{showcaptionsetup}{}{%
 \PassOptionsToPackage{caption=false}{subfig}}
\usepackage{subfig}
\makeatother

\usepackage{babel}

\usepackage[hidelinks]{hyperref}
\hypersetup{
    colorlinks=false,
}

\makeatletter
\def\ps@pprintTitle{%
  \let\@oddhead\@empty
  \let\@evenhead\@empty
  \let\@oddfoot\@empty
  \let\@evenfoot\@oddfoot
}
\makeatother
\journal{XYZ}

\begin{document}
\begin{frontmatter}
\title{Magnetoelastic Interactions Reduce Hysteresis in Soft Magnets}

\author[inst2]{Hongyi Guan}
\author[inst1]{Negar Ahani}
\author[inst3,inst4]{Carlos J. Garc\'ia-Cervera}
\author[inst2]{Ananya Renuka Balakrishna\corref{cor1}}
\ead{ananyarb@ucsb.edu}
\cortext[cor1]{Corresponding author}
\affiliation[inst2]{organization={Materials Department},
            addressline={University of California, Santa Barbara}, 
            city={Santa Barbara},
            postcode={93106}, 
            state={CA},
            country={USA}}
\affiliation[inst1]{organization={Aerospace and Mechanical Engineering},
            addressline={University of Southern California}, 
            city={Los Angeles},
            postcode={90089}, 
            state={CA},
            country={USA}}
\affiliation[inst3]{organization={Mathematics Department},
            addressline={University of California, Santa Barbara}, 
            city={Santa Barbara},
            postcode={93106}, 
            state={CA},
            country={USA}}
\affiliation[inst4]{organization={BCAM},
            addressline={Basque Center for Applied Mathematics}, 
            city={Bilbao},
            postcode={E48009}, 
            state={Basque Country},
            country={Spain}}

\begin{abstract}
The width of the magnetic hysteresis loop is often correlated with the material's magnetocrystalline anisotropy constant $\kappa_1$. Traditionally, a common approach to reduce the hysteresis width has been to develop alloys with $\kappa_1$ as close to zero as possible. However, contrary to this widely accepted view, we present evidence that magnetoelastic interactions governed by magnetostriction constants, elastic stiffness, and applied stresses play an important role in reducing magnetic hysteresis width, despite large $\kappa_1$ values. We use a nonlinear micromagnetics framework to systematically investigate the interplay between---material constants $\lambda_{100}, c_{11}, c_{12}$, $\kappa_1$, applied or residual stresses $\sigma_{\mathrm{R}}$, and needle domains---to collectively lower the energy barrier for magnetization reversal. A distinguishing feature of our work is that we correlate the energy barrier governing the growth of needle domains with the width of the hysteresis loop. This energy barrier approach enables us to capture the nuanced interplay between anisotropy constant, magnetostriction, and applied stresses, and their combined influence on magnetic hysteresis. We propose a mathematical relationship on the coercivity map: $\kappa_1 = \alpha(c_{11}-c_{12})(\lambda_{100}+\beta\sigma_{11})^2$ for which magnetic hysteresis can be minimized for an uniaxial residual stress $\sigma_\mathrm{R} = \sigma_{11}\hat{\mathbf{e}}_1\otimes\hat{\mathbf{e}}_1$ (and for some constants $\alpha,\beta$). These results serve as quantitative guidelines to design magnetic alloys with small hysteresis, and potentially guide the discovery of a new generation of soft magnets located beyond the $\kappa_1 \to 0$ region.
\end{abstract}


\end{frontmatter}

%Potential reviewers: 
%Dennis Kochmann, Bjorn Kiefer, Chad Landis, Yayoi Takamura, Kaushik Dayal, Vivekanand Dabade, Patrick Schamberger

\section{Introduction}

Ferromagnets exhibit a lag in magnetic response when cycled under an external field. This characteristic lag is called magnetic hysteresis and is a widespread phenomenon observed in ferromagnets. This hysteresis, on the one hand, is crucial to the design of memory-based devices, but on the other hand, signifies the energy loss during cycling \cite{silveyra2018soft, gutfleisch2011magnetic}. Despite the common occurrence of hysteresis across magnetic alloys, we neither understand its origins nor can we accurately predict its nonlinear manifestation in material systems \cite{bertotti2005science, james2015magnetic}. This gap in knowledge limits our ability to design and discover the next generation of magnetic alloys with minimal hysteresis.

One explanation for magnetic hysteresis is that the domain walls, which move back and forth in a magnetic alloy during cycling, get pinned to grain boundaries and other defects \cite{Foggiatto2022,jiles1986theory}. Kersten's ferromagnetic theory \cite{kersten1943underlying} and Becker's analysis of inclusion defects \cite{Neel1981} postulate that domain walls minimize surface energy by pinning to defects. These defects inhibit boundary movement and affect magnetization reversal. For example, today, non-magnetic precipitates with a large aspect ratio are synthesized to reduce domain wall pinning and contribute to lower hysteresis \cite{han2023strong}. However, the domain wall pinning theory does not account for the significant effect of the demagnetization field around defects \cite{Neel1981}.
%\footnote{\textcolor{green}{COMMENT--An idea for another time: These precipitates collectively called the Widmanst{\"a}tten patterns comprise thin and long rod-like inclusions that are oriented in the parent matrix at specific angles. See microstructure in \cite{han2023strong}. For a future work, it might be worth checking whether and how compatibility conditions between the precipitate the parent domain affects magnetization reversibility. The hypothesis would be --- if the Widmanst{\"a}tten precipitates lie along the compatible planes (satisfying hadarmard jump condition) the external field necessary to induce magnetization reversal will be small...}}

More recently, researchers have been developing modeling tools, in which machine learning methods are used to construct a continuum energy landscape \cite{Foggiatto2022,Kunii2022,schaffer2021machine}. In these studies, researchers employ feature extraction and dimensionality reduction techniques, such as principal component analysis, to investigate the effect of defect density, shape, and position on domain wall pinning. These studies provide insights into the interplay between magnetization and microstructural features in the alloy, and emphasize the role of morphological and structural defects in computing magnetic hysteresis. \textcolor{black}{Interestingly, however, bulk magnetic alloys such as the permalloy Fe$_{21.5}$Ni$_{78.5}$---synthesized independently by different research groups and likely with varying microstructures---consistently show similar magnetic hysteresis responses \cite{bozorth1993ferromagnetism, bozorth1953permalloy}. This uniformity in hysteresis behavior is surprising, given the typical variability in morphological features of an alloy from different synthesis conditions.} Furthermore, magnetic hysteresis persists in single-crystal defect-free magnetic alloys \cite{hubert1998magnetic}. These observations motivate us to question whether magnetic properties such as hysteresis is only a function of the microstructural features, or also affected by other factors including material constants, particle geometry, and processing techniques \cite{wu2024effect, mchenry1999amorphous, bozorth1993ferromagnetism}. % that is shaped by the magnetic domain evolution pathways and/or interaction with defects.

\begin{figure}
\begin{centering}
\includegraphics[width=\textwidth]{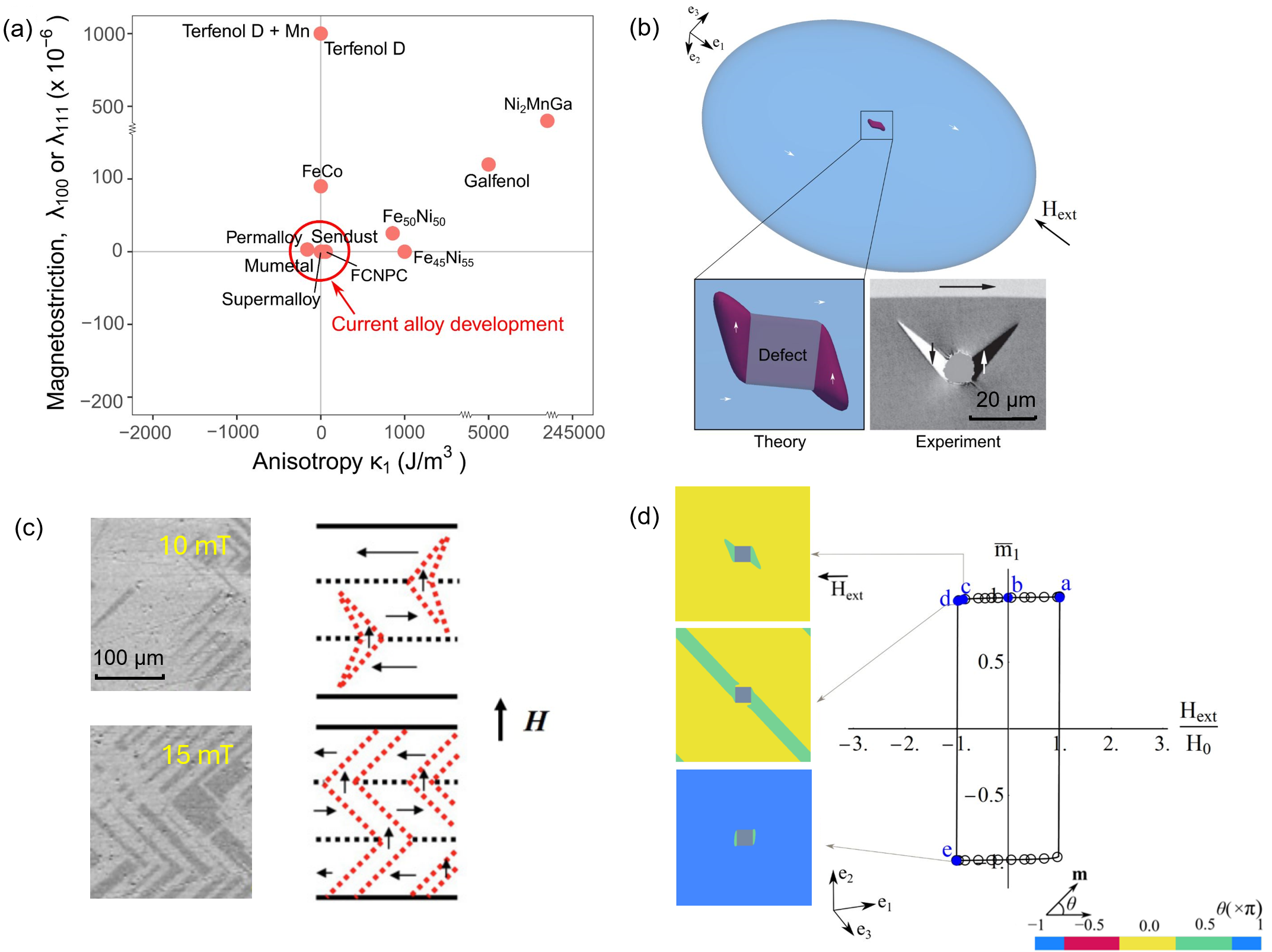}
\par\end{centering}
\caption{\label{Introduction}(a) The magnetocrystalline anisotropy and magnetostriction constants of commonly used soft magnetic alloys. Most soft magnetic alloys developed today have $\kappa_1 \to 0$, $\lambda_{100} \to 0$ and/or $\lambda_{111} \to 0$. However, outlier compounds, such as Fe$_{50}$Ni$_{50}$ with  $\kappa_1 \approx 1000$ $\mathrm{J/m^3}$ or Galfenol with $\kappa_1 \approx 5000$ $\mathrm{J/m^3}$ \cite{bozorth1993ferromagnetism}, highlight open questions on the role of other magnetic material constants on coercivity. (b) In our coercivity tool, we introduce a localized disturbance, in the form of a needle-domain, in a large magnetic ellipsoid body. These needle domains are commonly observed in experiments \cite{schafer2020tomography} and we theorize that these domains grow abruptly during magnetization reversal. The micrographs of the needle domain in iron-film are adapted with permission from \cite{schafer2020tomography} Copyright 2020 American Physical Society. (c) Micrographs and schematics showing the growth of needle domains into stripe domains from the nuclei at the 180$^{\circ}$ domain walls. This figure is adapted from Fig.~5(e) in Ref.~\cite{PhysRevMaterials.2.014412} with permission. Copyright 2018 American Physical Society. (d) A representative calculation of magnetic hysteresis using our coercivity tool. The needle domain grows in size under an external field and at a critical value, known as coercivity, the magnetization in the ellipsoid domain reverses. The inset figures show microstructural evolution during magnetization reversal, and the color bar represents the orientation of magnetization.}
\vspace{-5mm}
\end{figure}

Another explanation for magnetic hysteresis is related to the alloy's magnetocrystalline anisotropy constant $\kappa_1$. This anisotropy constant is a fundamental material constant that penalizes magnetization rotation away from the material's crystallographic easy axes and is thought to play a dominant role in magnetization reversal \cite{bozorth1993ferromagnetism,Mitra_2024}. That is, alloys with large $\kappa_1$ are typically characterized by large magnetic hysteresis and therefore a common strategy to develop magnetic alloys with small hysteresis is to systematically tune the alloy composition to reduce the anisotropy constant to zero $\kappa_1 \to 0$. Consequently, most of the soft magnets developed today are concentrated in the narrow material parameter space with $|\kappa_1| < 200$ J/m$^{3}$, see Fig.~\ref{Introduction}(a). However, there are a few outlier compounds that have small hysteresis despite their large $\kappa_1$ values, see Fig.~\ref{Introduction}(a) \cite{Rafique2004magnetic, Clark2003, Clemente2017, TICKLE1999627, Hu2001Magnetic}.\footnote{\textcolor{black}{Specimen geometries and the associated demagnetization fields significantly impact the magnetic hysteresis loops. For a given geometry, however, magnetic hysteresis can be further reduced by carefully tuning the alloy composition \cite{bozorth1993ferromagnetism}.}} These examples illustrate another gap in our understanding of the origins of magnetic hysteresis---that is, whether the anisotropy constant $\kappa_1$ is the only material constant governing the width of magnetic hysteresis.

Other magnetic material constants (e.g., magnetostriction, elastic stiffness, shape anisotropy constants) and mechanical constraints (e.g., residual stresses, substrate strains) are theorized to affect magnetization reversal \cite{james1998magnetostriction, miehe2011incremental, brown1966magnetoelastic}, but their precise role on coercivity is not well understood. For example, the magnetostriction constant---that describes the coupling between lattice strain and magnetization---is shown to hinder magnetization reversal and contributes to the power loss in soft magnetic cores \cite{tsukahara2022role}. This constant is typically on the order of a few micro-strains and is often neglected in magnetic hysteresis calculations \cite{wang2001gauss}. However, in magnetic alloys with large magnetostriction (e.g., Fe-Ga alloys) the interplay between elasticity and magnetism plays a crucial role in generating rich microstructural patterns and in governing the macroscopic deformation of the magnetic material \cite{chopra2015non, dabade2019micromagnetics}. Moreover, the presence of residual stresses and/or epitaxial strains affects magnetization reversal and is shown to alter the shape and size of hysteresis loops \cite{yang2021recent,kubota2012stress,ciubotariu2019strain}. These studies illustrate the prevalent role of magnetoelastic interactions on magnetization reversal. However, we do not clearly understand whether and how these interactions affect magnetic hysteresis and whether these interactions can be optimally designed to reduce magnetic coercivities \cite{lewis1964permalloy, becker1939ferromagnetismus, kittel1956ferromagnetic}.

In recent work, we showed the subtle but important role of magnetostriction on magnetic hysteresis \cite{balakrishna2021tool, renuka2022design, ARB2021Permalloy}. Using our newly developed coercivity tool \cite{balakrishna2021tool}, based on micromagnetics \cite{brown1963micromagnetics} and non-linear stability analysis, we demonstrated that the energy barrier for magnetization reversal is related to a delicate interplay between anisotropy $\kappa_1$ and magnetostriction $\lambda_{100}$ constants. This interplay between magnetic constants has been overlooked in previous works because of the extremely small values of magnetostriction constants (on the order of $\sim10^{-6}$ strains) and, the mathematical models used to investigate hysteresis are based on the linear approximation of the free energy function that often overestimates magnetic hysteresis  \cite{brown1963micromagnetics}. These methods miss the shoulder of hysteresis loops and are not suitable to elucidate the fine balance between material constants.\footnote{This discrepancy in coercivity values between theoretical predictions and experimental measurements is called Brown's `coercivity paradox' \textcolor{black}{\cite{brown1963micromagnetics, brown1966magnetoelastic,PhysRevB.105.214431}}.} In contrast to these earlier works, our computations provide insights into the longstanding permalloy problem (i.e., why the specific composition of the $\mathrm{Fe_{21.5}Ni_{78.5}}$ alloy has the smallest hysteresis) \cite{ARB2021Permalloy}, and quantitatively establishes the role of the magnetostriction constant $\lambda_{100}$ on lowering coercivity at the 78.5\% Ni-composition. Furthermore, we propose a parabolic relation between material constants $\frac{(c_{11}-c_{12})\lambda_{100}^2}{\kappa_1} = \mathrm{Const.}$, for which magnets can have small hysteresis \cite{renuka2022design}. These results initiate a new line of research, in which multiple material parameters can be systematically designed to reduce coercivity in soft magnets \cite{renuka2022design}.

The magnetostriction $\lambda_{100}$, however, is one such magnetoelastic constant, and the effect of other material properties---most notably elastic stiffness, and residual stresses---on magnetization reversal is unknown. For example, the elastic stiffness constants $\mathbb{C} (c_{11}, c_{12}, c_{44})$ determine the deformation of a magnet under an applied mechanical/magnetic load and contributes to the micromagnetic energy terms governing magnetization reversal. While $c_{11}, c_{12}$ affect magnetization reversibility, their precise role on hysteresis loops and coercivities is not well understood. Moreover, heat treatment of soft magnets and several applications of soft magnets (e.g., memory elements, sensors) introduce residual stresses (and/or thermal stresses) $\sigma_{\mathrm{R}}$ in the material \cite{Panduranga2021}. These pre-existing stress states in the material coupled with their intrinsic stiffness constants are shown to alter fundamental magnetic properties and therefore affect magnetization reversal \cite{meisenheimer2021engineering, chakrabarti2022magnetic, hsieh2021strain}. We hypothesize that these magnetoelastic interactions affect magnetic hysteresis and can be systematically designed to dramatically reduce coercivity.

In this paper, we test this hypothesis using our nonlinear micromagnetics framework \cite{balakrishna2021tool, ARB2021Permalloy, renuka2022design}. We use the magnetic alloy Fe$_{50}$Ni$_{50}$ as a representative example\footnote{The Fe$_{50}$Ni$_{50}$ alloy has a large anisotropy constant $\kappa_1 \approx 1000$ $\mathrm{J/m^3}$ and is an example of a magnet with relatively large coercivity (e.g., in comparison to the permalloy Fe$_{21.5}$Ni$_{78.5}$). This alloy serves as a suitable example of a `bad' soft magnet with large coercivity, in which the magnetoelastic interactions can be designed to reduce its hysteresis, and serves as a representative example in this work.}, and investigate whether and how material constants, such as magnetostriction ($\lambda_{100}$), elastic stiffness ($c_{11}, c_{12}$), and residual stresses $\sigma_{\mathrm{R}}$, contribute to hysteresis width, and how we can design these magnetoelastic interactions to reduce magnetic hysteresis (Study 1). We next systematically scan the material parameter space (by computing $N = 726$ independent micromagnetic calculations) to identify suitable combinations of residual stresses, magnetostriction, and elastic stiffness, for which magnetic hysteresis can be dramatically reduced (Study 2). Our key finding is that for a specific combination of magnetoelastic constants, $\lambda_{100}$, $c_{11}$, $c_{12}$, and $\sigma_{\mathrm{R}}$, the height of energy barrier for magnetization reversal is lowered. We theoretically analyze the energy barrier governing magnetization reversal and find that coercivity is minimum when the magnetoelastic constants approximately satisfy the parabolic relationship $\kappa_1 = \alpha(c_{11}-c_{12})(\lambda_{100}+\beta\sigma_{11})^2$. Our analysis, based on the concept of energy barriers, explains the mathematical relationship between material constants and applied or residual stress in lowering magnetic hysteresis. These results establish coercivity maps that provide quantitative insights into how elastic stiffness constants $\mathbb{C}$, magnetostriction constants $\lambda_{100}$, and residual stresses $\sigma_{\mathrm{R}}$ affect the parabolic locus for minimum coercivities. These findings would serve as design guidelines to systematically develop magnetic alloys with small hysteresis and enable the discovery of a new generation of soft magnets that are located beyond the $\kappa_1 \to 0$ region in Fig.~\ref{Introduction}(a).

\section{Methods}

Traditionally, researchers compute magnetic hysteresis by doing linear stability analysis of a uniformly magnetized domain \cite{brown1963micromagnetics}. In this technique, a second variation of the micromagnetics energy is simplified to a eigenvalue problem and its non-trivial solutions are associated with the magnetization reversal. A well known drawback of this approach is that, when the material constants are substituted (e.g., of iron) the solutions overestimate hysteresis widths by several orders of magnitude. This discrepancy between the theoretical prediction and experimental measurement of the coercive force is called the `coercivity paradox' \cite{brown1963micromagnetics, brown1966magnetoelastic}. Furthermore, many of the standard techniques to compute magnetic hysteresis do not account for magnetostriction---a reasonable assumption given that this material constant is often small (on the order of $\sim \mu\epsilon = 10^{-6}$) in well known soft magnetic alloys.

In a departure from these earlier methods to predict coercivity, we compute magnetic hysteresis in terms of energy barriers \cite{https://doi.org/10.1002/zamm.201800179, ZHANG20094332, CHEN20132566, Cui2006} and localized disturbances \cite{balakrishna2021tool, ARB2021Permalloy, renuka2022design}. In this technique, we consider a pre-existing nucleus, which serves as a large localized disturbance to an otherwise uniformly magnetized domain, and analyze the energy barriers associated with the nucleus growth. By lowering the energy barrier we arrive at a mathematical relation between material constants for which hysteresis is reduced. This approach has been successfully used to reduce thermal hysteresis in shape-memory alloys \cite{knupfer2013nucleation, zhang2009energy} (another type of phase transformation material) and presents a way forward to designing magnetic hysteresis.

For example, in our recent works on soft magnetic alloys \cite{balakrishna2021tool, ARB2021Permalloy, renuka2022design, balakrishna2022compatible}, we consider nuclei in the form of needle domains that are commonly observed around defects and/or at sharp corners. These needle domains form to minimize the demagnetization energy, see Fig.~\ref{Introduction}(b) \cite{hubert1998magnetic}. Under an external field, these needle domains are the first to grow in size (first in length and then in width), leading to a complete magnetization reversal \cite{hubert1998magnetic,balakrishna2021tool}. In our modeling framework, we treat these needle domains as non-negligible localized disturbances and correlate their growth with the width of the hysteresis loop. By doing so, we trace the hysteresis loop for a soft magnet and predict its coercivity, see Fig.~\ref{Introduction}(d).\footnote{This approach of computing magnetic hysteresis has been applied for the Fe-Ni alloy system and explains the small coercivity at the permalloy composition ($\mathrm{Fe_{21.5}Ni_{78.5}}$) \cite{ARB2021Permalloy}. The coercivity tool also provides insights into the energy barriers, arising from magnetostriction, which affect hysteresis in soft magnetic alloys \cite{balakrishna2021tool, renuka2022design}.} In this work, we build on this technique to not only derive a relation between material constants but also consider the role of applied stresses in reducing magnetic hysteresis. %\textcolor{blue}{\sout{We use this modeling framework to investigate the effect of magnetoelastic interactions on magnetic hysteresis.}}

Our coercivity tool is based on micromagnetics with magnetostriction \cite{brown1963micromagnetics, brown1966magnetoelastic}. The total free energy $\Psi$ for a magnetic body $\mathcal{E} \subset \mathbb{R}^3$ is a functional of magnetization $\mathbf{m} = \frac{\mathbf{M}}{m_s} = \mathrm{m}_1{\hat{\mathbf{e}}}_1 + \mathrm{m}_2{\hat{\mathbf{e}}}_2 + \mathrm{m}_3{\hat{\mathbf{e}}}_3$ and strain $\mathbf{E}$:
\begin{align}
{\Psi} & =\int_{\mathcal{E}}\Bigg\{\nabla\mathbf{m}\cdot\mathbf{A}\nabla\mathbf{m}+\kappa_{1}(\mathrm{m}_{1}^{2}\mathrm{m}_{2}^{2}+\mathrm{m}_{2}^{2}\mathrm{m}_{3}^{2}+\mathrm{m}_{3}^{2}\mathrm{m}_{1}^{2})+\frac{1}{2}[\mathbf{E}-\vb{E}_0(\vb{m})]\cdot\mathbb{C}[\mathbf{E}-\vb{E}_0(\vb{m})]\nonumber\\
&  -\mu_{0}m_s\mathbf{H_{\mathrm{ext}}\cdot m} - \mathbf{\sigma}_{\mathrm{R}}\cdot\mathbf{E}\Bigg\}\mathrm{d\mathbf{x}} +\int_{\mathbb{R}^{3}}\frac{\mu_{0}}{2}\left|\mathbf{H}_\mathrm{d}\right|^{2}\mathrm{d\mathbf{x}}.\label{MicromagneticsEnergy}
 \end{align}
In Eq.~\ref{MicromagneticsEnergy}, the spontaneous strain $\vb{E}_0(\vb{m})$ for the cubic crystal is
\begin{equation}
    \vb{E}_0(\vb{m}) = \dfrac{3}{2}\mqty(\lambda_{100}\mathrm{(m_1^2-1/3)} & \mathrm{\lambda_{111}m_1m_2} & \mathrm{\lambda_{111}m_1m_3} \\ \mathrm{\lambda_{111}m_1m_2} & \lambda_{100}\mathrm{(m_2^2-1/3)} & \mathrm{\lambda_{111}m_2m_3} \\ \mathrm{\lambda_{111}m_1m_3} & \mathrm{\lambda_{111}m_2m_3} & \lambda_{100}\mathrm{(m_3^2-1/3)}).
\end{equation}
In Eq.~\ref{MicromagneticsEnergy}, $\mathbf{H}_{\mathrm{ext}}$ corresponds to the applied magnetic field and $\mathbf{\sigma}_{\mathrm{R}}$ corresponds to the applied mechanical stress (or residual stress). The fundamental material constants, such as the gradient energy constant $\mathrm{A}$, anisotropy constant $\kappa_1$, elastic stiffness constant $\mathbb{C}$, magnetostriction constants $\lambda_{100}, \lambda_{111}$, and the saturation magnetization $m_s$ determine the free energy landscape. 
%The specific values of the magnetic material constants determine the energy barrier for magnetization reversal and are discussed in detail in the supplement (see Table.~\ref{Symbols}-\ref{tab:materialconstant}) and in Refs.~\cite{balakrishna2021tool, renuka2022design}. 
Further details of the notations in Eq.~\ref{MicromagneticsEnergy} are discussed in the supplementary information and in Refs.~\cite{balakrishna2021tool, renuka2022design}.

We \textcolor{black}{describe the magnetization dynamics} based on the generalized Landau-Lifshitz-Gilbert equation Eq.~\ref{LLG}, and solve for magnetostatic and mechanical equilibrium conditions Eq.~\ref{eq:magnetostaticeq}--\ref{eq:mechanicaleq}:
\begin{gather}
\frac{\partial\mathbf{m}}{\partial \tau}=-\mathbf{m}\times\mathcal{H}-\alpha\mathbf{m}\times(\mathbf{m}\times\mathcal{H}) \label{LLG}\\
\nabla\cdot(\mathbf{\mathbf{-\nabla\zeta_{\mathbf{m}}}+\mathit{m_s}m})  =0\qquad\text{in }\mathbb{R}^{3}\label{eq:magnetostaticeq}\\
\nabla\cdot\mathbf{\sigma}  = -\nabla\cdot\mathbb{C}[\vb{E}-\vb{E}_0(\vb{m})]= \vb{0}\qquad\text{in }\mathcal{E}.\label{eq:mechanicaleq}
\end{gather}
Here, $\mathcal{H}=-\frac{1}{\mu_0 m_s^2}\frac{\delta\Psi}{\delta\mathbf{m}}$
is the effective field, $\tau=\gamma m_s t$ is the dimensionless time step, $\gamma$ is the gyromagnetic ratio, and $\alpha$ is the damping constant. We numerically solve Eqs.~\ref{LLG}--\ref{eq:mechanicaleq} in Fourier space using the Gauss-Seidel projection method \cite{wang2001gauss}. Further details on the numerics can be found in Refs. \cite{wang2001gauss,balakrishna2021tool,zhang2005phase}. In this study, we calibrate our micromagnetics model for the Fe$_{50}$Ni$_{50}$ alloy, see Tables~\ref{Symbols}--\ref{tab:materialconstant} for material constants.

To compute coercivity, we model a 3D domain $\Omega$ of size $128\times128\times22$ grid points with a nonmagnetic inclusion $\Omega_d$ of size $8\times8\times6$ at its geometric center. This 3D domain is embedded within a large plate-like smooth ellipsoid $\mathcal{E}$ and the demagnetization factor associated with this oblate ellipsoid is $\vb{N}_\mathrm{d} = \hat{\vb{e}}_3\otimes \hat{\vb{e}}_3$), see Fig.~\ref{Introduction}(b). In our model, we use the ellipsoid and reciprocal theorems to compute the demagnetization fields arising from the poles of the ellipsoid geometry, see the Supplement in Ref.~\cite{balakrishna2021tool}.  That is, we estimate the coercivity of a bulk magnetic ellipsoid by numerically modeling the 3D domain with suitable boundary conditions (see Ref.~\cite{balakrishna2021tool} for details). We initialize the 3D domain with a uniform magnetization $\mathrm{\mathbf{m}=\hat{\mathbf{e}}_1},$ (except at the defect site, where $\mathbf{m=0}$) along the material's easy axes. A needle-like domain forms around the defect to minimize the demagnetization energy. We apply an external field $\vb{H}_{\mathrm{ext}}$ parallel to the initial magnetization, which is large enough to saturate it, see Fig.~\ref{Introduction}(b). We then decrease this applied field and the needle domain around the defect grows until a critical field is reached at which the magnetization reverses abruptly. We trace the hysteresis loop and quantify coercivity for each combination of elastic stiffness constants ($c_{11}$, $c_{12}$) and residual stresses $\sigma_{\mathrm{R}}$.

\section{Results}\label{sec:Results}
%1000 words
\subsection*{Magnetoelastic Interactions and Hysteresis}
\begin{figure}[t]
\begin{centering}
\textcolor{black}{\includegraphics[width=\textwidth]{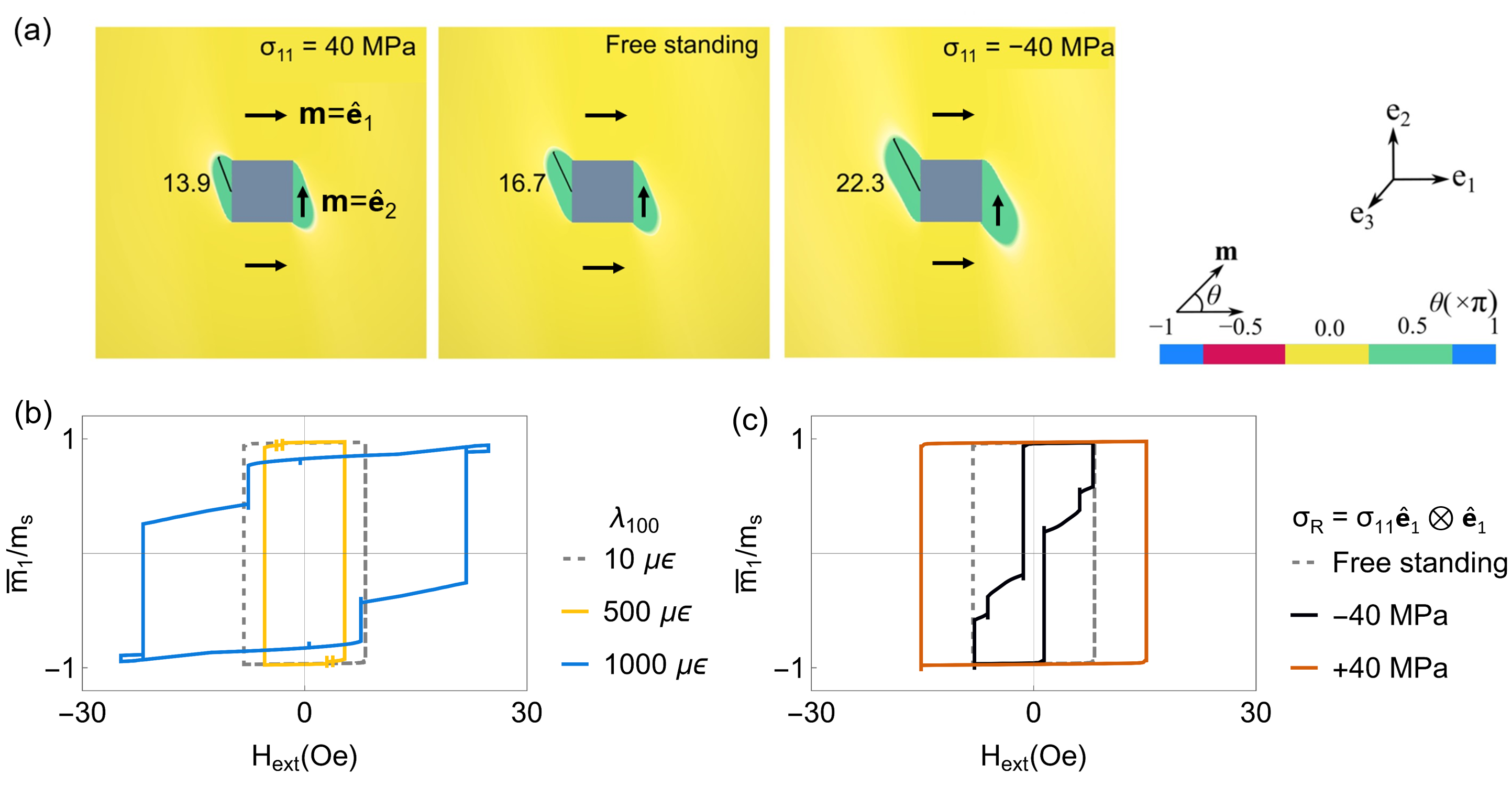}}
\par\end{centering}
\caption{\label{CoercivityPlot}(a) The effect of residual stresses $\sigma_{\mathrm{R}}=\sigma_{11}\hat{\mathbf{e}}_1\otimes\hat{\mathbf{e}}_1$ affect the length of the needle domain microstructures. For example, compressive stresses ($\approx -40$ $\mathrm{MPa}$) stabilize longer needle domains, making them more susceptible to abrupt growth under an external magnetic field. (b) magnetostriction $\lambda_{100}$ and (c) residual stress $\sigma_{11}$ on the shape and size of hysteresis loops. The gray-dashed line in subfigures (b-c) corresponds to the Fe$_{50}$Ni$_{50}$ reference alloy.}
\end{figure}
%\textcolor{red}{Thanks for the update, Hongyi. In subfigure (a) can you make the arrows of similar size and maybe add $\mathbf{m}$ near one of the arrows (in any of the subfigures) to indicate what they represent. And, for subfigure (c) can you write $\sigma_{\mathrm{R}} = \sigma_{11}\hat{\mathbf{e}}_1\otimes\hat{\mathbf{e}}_1$}}

In Study 1, we test our hypothesis that magnetoelastic interactions arising from magnetostriction, elastic stiffness, and residual stresses can be optimally designed to lower coercivity in soft magnetic alloys despite the alloy's large magnetocrystalline anisotropy. As a reference case, we investigate the effect of magnetoelastic interactions on the hysteresis loop of a soft magnetic alloy Fe$_{50}$Ni$_{50}$. This alloy is known to have a larger coercivity than the permalloy Fe$_{21.5}$Ni$_{78.5}$ because of its relatively large magnetocrystalline anisotropy constant, $\kappa_1 \approx 1000$ $\mathrm{J/m^3}$. In this study, we use Fe$_{50}$Ni$_{50}$ as a candidate material and show that we can further reduce its coercivity by suitably designing its magnetoelastic interactions. By doing so, we demonstrate a potential material design approach in which magnetostriction, residual stresses, and elastic stiffness constants are systematically designed to reduce coercivity. This design strategy follows our previously proposed mathematical design rule \cite{renuka2022design}, i.e., hysteresis in soft magnetic alloys is minimized close to the $\frac{(c_{11}-c_{12})\lambda_{100}^2}{2 \kappa_1} = $ Const. parabola.\footnote{The parabolic relation derived in our previous work \cite{renuka2022design}, suggests that in addition to achieving minimum coercivity for $\kappa_1 \to 0$ there exists other regions along $\frac{(c_{11}-c_{12})\lambda_{100}^2}{2 \kappa_1}$ = Const. at which coercivity is small. While this relation is based on $\langle 100 \rangle$ easy axes and might differ for alloys with different easy axes (and/or in the presence of residual stresses). The parabolic relation is intended to be a guide to the search for soft magnetic alloys with large $\kappa_1$.}

At the microstructural level, we note that the compressive \textcolor{black}{($\sigma_{11} < 0$)} and tensile \textcolor{black}{($\sigma_{11} > 0$)} stresses have opposing effects on the needle domain geometry, see Fig.~\ref{CoercivityPlot}(a). \textcolor{black}{On the one hand, compressive stresses $\sigma_{11} < 0$, applied to alloys with $\lambda_{100} > 0$, energetically favors magnetization along the $\hat{\mathbf{e}}_2$ directions, i.e., $\mathbf{m}=\pm\hat{\mathbf{e}}_2$. These stresses when applied to a free-standing microstructure with needle domains leads to longer needles in the equilibrium state.} %These compressive stresses favor lattice distort  the magnetization within the needle domain ($\vb{m} \approx (0,1,0)$, with spontaneous strain $0$ at $\hat{\vb{e}}_{1}$ direction) is more energetically favorable than the magnetization outside it ($\vb{m} \approx (1,0,0)$, with spontaneous strain $\lambda_{100}>0$ at $\hat{\vb{e}}_{1}$ direction), leading to longer needles when compared to the free-standing microstructural state.
In our calculations, these longer needles require a smaller external field for domain growth and magnetization reversal and $\sigma_{11} < 0$ lowers the coercivity of the magnetic alloy, see Fig.~\ref{CoercivityPlot}(c). On the other hand, tensile stresses shorten the length of the needle domains, and these shorter needles require a relatively stronger external field for complete magnetization reversal. These results indicate that applied stresses can have a significant effect on the magnetic properties of materials, and the magnitude and type of the applied stresses can have opposite effects on coercivity.

Fig.~\ref{CoercivityPlot}(b-c) illustrates the effect of two different factors, namely magnetostriction and residual stresses, on the magnetic hysteresis of a Fe$_{50}$Ni$_{50}$. Fig.~\ref{CoercivityPlot}(b) shows that increasing the value of magnetostriction $\lambda_{100}$ from $10$ $\mu\epsilon$ to $500$ $\mu\epsilon$ (with 1 $\mu\epsilon = 10^{-6}$) reduces the coercivity from $8.16$ $\mathrm{Oe}$ to $5.3$ $\mathrm{Oe}$. However, further increasing the value of magnetostriction from 500 $\mu\epsilon$ to 1500 $\mu\epsilon$ results in a four-fold increase in coercivity, from 5.3~$\mathrm{Oe}$ to 21.73~$\mathrm{Oe}$. This shows that, despite the large magnetocrystalline anisotropy constant of the alloy, there is an optimal value of magnetostriction (i.e., approximately satisfying $\frac{(c_{11}-c_{12})\lambda_{100}^2}{2 \kappa_1} = $ Const.) that minimizes coercivity. Similarly, Fig.~\ref{CoercivityPlot}(c) shows that applying compressive residual stresses of $\sigma_{11}=-40$ $\mathrm{MPa}$ to the Fe$_{50}$Ni$_{50}$ alloy dramatically reduces its coercivity from $8.16$ $\mathrm{Oe}$ to $1.41$ $\mathrm{Oe}$.\footnote{These residual stresses correspond to an uniaxial strain of about $\pm200$ $\mu\epsilon$ (with typical Young's Modulus of about $200$ $\mathrm{GPa}$) for soft magnetic alloys. As noted before, these residual stresses could naturally arise from mechanical constraints or defects during heat treatment and are similar in value to the thermal strains in soft magnets \cite{li2016giant}.} However, when tensile stresses of similar magnitude are applied to the Fe$_{50}$Ni$_{50}$ alloy, the coercivity value increases to $15$ $\mathrm{Oe}$. 

Figs.~\ref{CoercivityPlot}(b-c) show that a change in magnetostriction or residual stresses not only affects coercivities but also affects the shape of hysteresis loops. For example, when $\lambda_{100}$ is increased from $500$ to $1500$ $\mu\epsilon$, pinched or step-like features emerge on the hysteresis loop, see Fig.~\ref{CoercivityPlot}(b). Similar features are observed in the hysteresis loop of Fe$_{50}$Ni$_{50}$ with residual stress of $\sigma_{11} = -40$ $\mathrm{MPa}$. These step-like features correspond to the new intermediate microstructural states. For example, $\lambda_{100} = 1500$ $\mu\epsilon$ in Fig.~\ref{CoercivityPlot}(b), magnetizations are approximately oriented along ($\sqrt{\frac{2}{3}},\sqrt{\frac{1}{3}},0$) and/or the ($\sqrt{\frac{1}{3}},\sqrt{\frac{2}{3}},0$) directions. These intermediate states with magnetization orientation away from the easy axes (i.e., $\langle 100 \rangle$ for alloys with $\kappa_1 > 0$) are supported by the relatively large magnetostriction constant ($\lambda_{100} = 1500$ $\mu\epsilon$) or the significant applied stress ($\sigma_{11} = -40$ MPa).

%\textcolor{blue}{\sout{In contrast to magnetostriction and residual loads, we find that varying the elastic stiffness coefficients in the range $(c_{11}-c_{12}) \in \{1$ $\mathrm{GPa}, 100$ $\mathrm{GPa}, 400$ $\mathrm{GPa}\}$ has a negligible effect on magnetic coercivities for the Fe$_{50}$Ni$_{50}$ alloy. This is because the magnetostriction constant of Fe$_{50}$Ni$_{50}$ is small $\lambda_{100} = 10$ $\mu\epsilon$ and increasing the stiffness constants---by two orders of magnitude---does not significantly affect the height of energy barriers governing magnetization reversal.}\footnote{This energy barrier is estimated at the shoulder of the hysteresis loop and accounts for contributions from the magnetocrystalline anisotropy, elastic energy, and demagnetization energy terms. The Zeeman energy acts like a driving force, which drives the system to cross the energy barrier.} \sout{We further explore the effect of elastic stiffness constants for a range of magnetostriction and residual stress values in Study 2.}}

Overall, Fig.~\ref{CoercivityPlot} shows that, soft magnets with suitable magnetostrictive coefficients can have small hysteresis despite the large $\kappa_1$ values. These magnetoelastic interactions could be achieved in an experimental setting through controlled alloy doping \cite{10.1063/1.3268471,10.1063/1.5007248,albertini2011,Meisenheimer2021}, additive manufacturing \cite{9186272,Yang2023}, or epitaxial growth of materials on substrates with appropriate pre-strains \cite{Chumak2021}. Furthermore, the tailoring of magnetoelastic interactions not only allows for the design of coercivities but also for the shape of hysteresis loops, which would be relevant for soft magnetic applications in actuators and memory devices.

\subsection*{Energy Barriers and Coercivity Maps}
In Study 2, we systematically scan the material parameter space (by computing $N = 726$ independent micromagnetic calculations) to identify suitable combinations of residual stress, magnetostriction, and elastic stiffness constants, for which magnetic hysteresis can be dramatically reduced. Specifically, we compute the coercivity of soft magnets within the following parameter ranges: elastic stiffness constants $90~\mathrm{GPa} \leq c_{11} \leq 239$ $\mathrm{GPa}$, magnetostriction constants $-1000~\mu\epsilon \leq \lambda_{100} \leq 1000~\mu\epsilon$ (with $\lambda_{111} = 0$) and anisotropy constants $0 \leq \kappa_1 \leq 2000$ $\mathrm{J/m^3}$. Furthermore, we study the effect of residual stresses $\sigma_\mathrm{R} = \sigma_{11}\hat{\mathbf{e}}_1\otimes\hat{\mathbf{e}}_1,$ $\sigma_{11} = \pm 200$ $\mathrm{kPa}$ on coercivity across the material parameter space. In these calculations, we analyze the energy contributions from magnetocrystalline anisotropy, magnetostriction, and applied loads, all of which collectively govern magnetization reversal. Overall, our computations establish coercivity maps for a range of magnetoelastic interactions characterized by material parameters such as $\sigma_{\mathrm{R}}$, $\mathbb{C}$, and $\lambda_{100}$, offering a quantitative guideline for alloy design and development.

\begin{figure}
\begin{centering}
\textcolor{black}{\includegraphics[width=\textwidth]{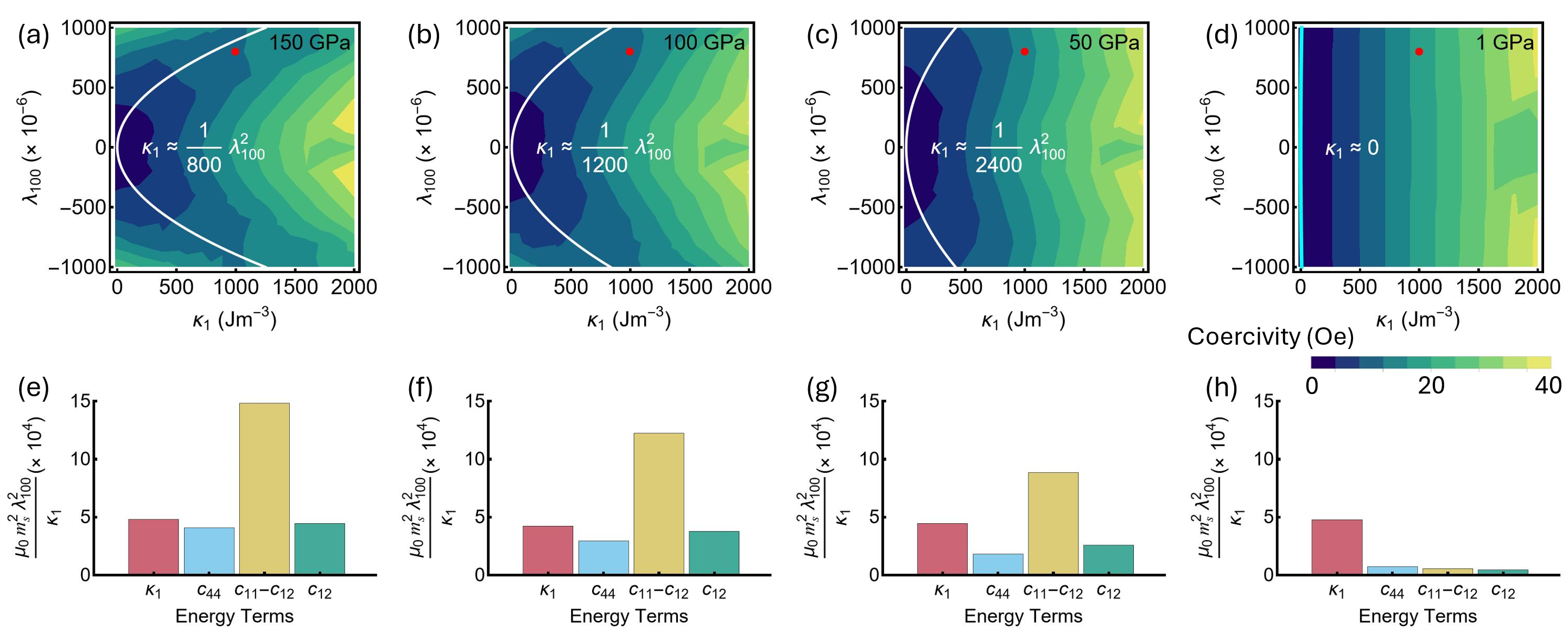}}
\caption{\label{CoercivityMaps} Contour plots showing coercivity values for each combination of magnetostriction constant $\lambda_{100}$ and anisotropy constant $\kappa_1$ for different elastic stiffness constants (a) $c_{11}-c_{12} = 150$ $\mathrm{GPa}$, (b) $c_{11}-c_{12} = 100$ $\mathrm{GPa}$, (c) $c_{11}-c_{12} = 50$ $\mathrm{GPa}$, and (d) $c_{11}-c_{12} = 1$ $\mathrm{GPa}$. The parabolic locus of minimum coercivity is shown in white on each of these contour plots. For clarity, the parabola for $c_{11}-c_{12} = 1$ $\mathrm{GPa}$ is highlighted in cyan. The histograms (e-h) show energy contributions at a representative point (shown by the red dot in (a-d)) with $\kappa_1 = 1000$ $\mathrm{J/m^3}$ and $\lambda_{100} = 800$ $\mu\epsilon$. The energies in these subfigures are evaluated in unit of $(\mu_0 m_s^2\lambda_{100}^2/\kappa_1)\times 10^{4}$. In subfigures (e-g), the primary elastic energy term contributing to the height of the energy barrier is the coupled magnetoelastic term with the multiplier $c_{11}-c_{12}$.}
\par\end{centering}
\end{figure}

Fig.~\ref{CoercivityMaps}(a-d) shows the coercivity maps for soft magnets as a function of magnetostriction $\lambda_{100}$, anisotropy $\kappa_1$, and elastic stiffness constants $c_{11}, c_{12}$. These maps show that magnetic alloys with large anisotropy constants $\kappa_1$ can have small coercivities that are comparable to alloys with $\kappa_1 \to 0$, provided the magnetostriction constants of these alloys satisfy a certain parabolic locus $\kappa_1 \propto \lambda_{100}^2$. Furthermore, Fig.~\ref{CoercivityMaps}(a-d) shows that the elastic stiffness $c_{11}-c_{12}$ constants govern the width of this parabolic locus. For example, magnets with large elastic stiffness $c_{11}-c_{12} = 150$ $\mathrm{GPa}$ have a narrower parabolic locus for minimum coercivity ($\kappa_1 \approx \frac{1}{800}\lambda_{100}^2$), than magnets with small elastic stiffness $c_{11}-c_{12} = 50$ $\mathrm{GPa}$ that have a wider parabolic locus for minimum coercivity ($\kappa_1 \approx \frac{1}{2400}\lambda_{100}^2$). In magnets with extremely small stiffness constants $c_{11}-c_{12} = 1$ $\mathrm{GPa}$, the parabolic locus for minimum coercivity reduces to be approximately a line along $\kappa_1 = 0$. 

To interpret the effect of stiffness constants on the parabolic locus, we analyze the dominant energy terms governing magnetization reversal for a representative alloy (with $\kappa_1 = 1000~\mathrm{J/m}^3$ and $\lambda_{100} = 800~\mu\epsilon$), see Figs.~\ref{CoercivityMaps}(e-h).\footnote{The energy contributions are analyzed on the shoulder of hysteresis loops, at which point the needle microstructure has grown to its full length and this transient microstructural state precedes the complete magnetization reversal in the computational domain, see Fig.~\ref{TransientState}(b).} We theoretically analyze these energy barriers in the next section, but for now we compare the different contributions from the anisotropy energy term (labelled as $\kappa_1$ in Figs.~\ref{CoercivityMaps}(e-h)):
\begin{equation}
    \int_{\mathcal{E}} \kappa_{1}(\mathrm{m}_{1}^{2}\mathrm{m}_{2}^{2}+\mathrm{m}_{2}^{2}\mathrm{m}_{3}^{2}+\mathrm{m}_{3}^{2}\mathrm{m}_{1}^{2}) \dd \vb{x}, \label{Aniso}
\end{equation}
and the three magneto-elastic energy terms (in the absence of applied mechanical loads): 
\begin{equation}
\begin{aligned}
    &\int_{\mathcal{E}} \frac{1}{2}[\mathbf{E}-\mathbf{E_{0}\mathrm{(\mathbf{m})}}]\cdot\mathbb{C}[\mathbf{E}-\mathbf{E_{0}\mathrm{(\mathbf{m})}}]  \dd \vb{x} \\
    =& \int_{\mathcal{E}}\Bigg\{ \underbrace{2c_{44}\sum_{\substack{i,j=1 \\ i>j}}^3\left(\varepsilon_{ij}-\frac{3}{2}\lambda_{111}\mathrm{m}_i\mathrm{m}_j\right)^2}_{\displaystyle{c_{44}\text{ term}}}+\underbrace{\frac{c_{11}-c_{12}}{2}\sum_{i=1}^3\qty[\varepsilon_{ii}-\frac{3}{2}\lambda_{100}\qty(\mathrm{m}_i^2-\frac{1}{3})]^2}_{\displaystyle{c_{11}-c_{12}\text{ term}}} \\ &+\underbrace{\frac{c_{12}}{2}\qty(\sum_{i=1}^3\varepsilon_{ii})^2}_{\displaystyle{c_{12}\text{ term}}}\Bigg\}\dd\vb{x}.
\end{aligned} \label{ElThreeTerms}
\end{equation}
These elastic energy terms are labeled as `$c_{44}$', `$c_{11}-c_{12}$', `$c_{12}$' in Figs.~\ref{CoercivityMaps}(e-h) and the strain tensors is written as:
\begin{equation}
    \vb{E} = \sum_{i,j}\varepsilon_{ij} (\vb{\hat{e}}_{i}\otimes\vb{\hat{e}}_{j}).
\end{equation}
%We find the micromagnetic energy contribution arising from the $(c_{11}-c_{12})$ term in Eq.~\ref{ElThreeTerms} to dominate magnetization reversal in elastically stiff magnets, i.e., $(c_{11}-c_{12}) > 50$ $\mathrm{GPa}$, and the anisotropy energy term to dominate magnetization reversal in magnets with $(c_{11}-c_{12}) \le 1$ $\mathrm{GPa}$. Thus, in magnetic alloys with small stiffness constants, the energy barrier arising from magnetoelastically coupled terms is negligible, and the anisotropy energy term dominates magnetization reversal.
We find the micromagnetics energy contribution arising from the $c_{11}-c_{12}$ term in Eq.~\ref{ElThreeTerms} to dominate magnetization reversal in magnets with $c_{11}-c_{12} \geq 50$ $\mathrm{GPa}$. However, for magnetic alloys with stiffness constants of $c_{11}-c_{12}$ = 1 GPa (see Fig.~\ref{CoercivityMaps}(d)), the elastic energy contribution is negligible and the anisotropy energy term dominates magnetization reversal. The axial strains in Eq.~\ref{ElThreeTerms} scale linearly with the magnetostriction constant $\varepsilon_{ii}\propto\lambda_{100}$ and the elastic energy approximately scales as $(c_{11}-c_{12})\lambda_{100}^2$. This elastic energy term, with suitable stiffness and magnetostriction constants, plays a significant role in reducing coercivity in Figs.~\ref{CoercivityMaps}(a-c). 

\begin{figure}
\begin{centering}
\textcolor{black}{\includegraphics[width=\textwidth]{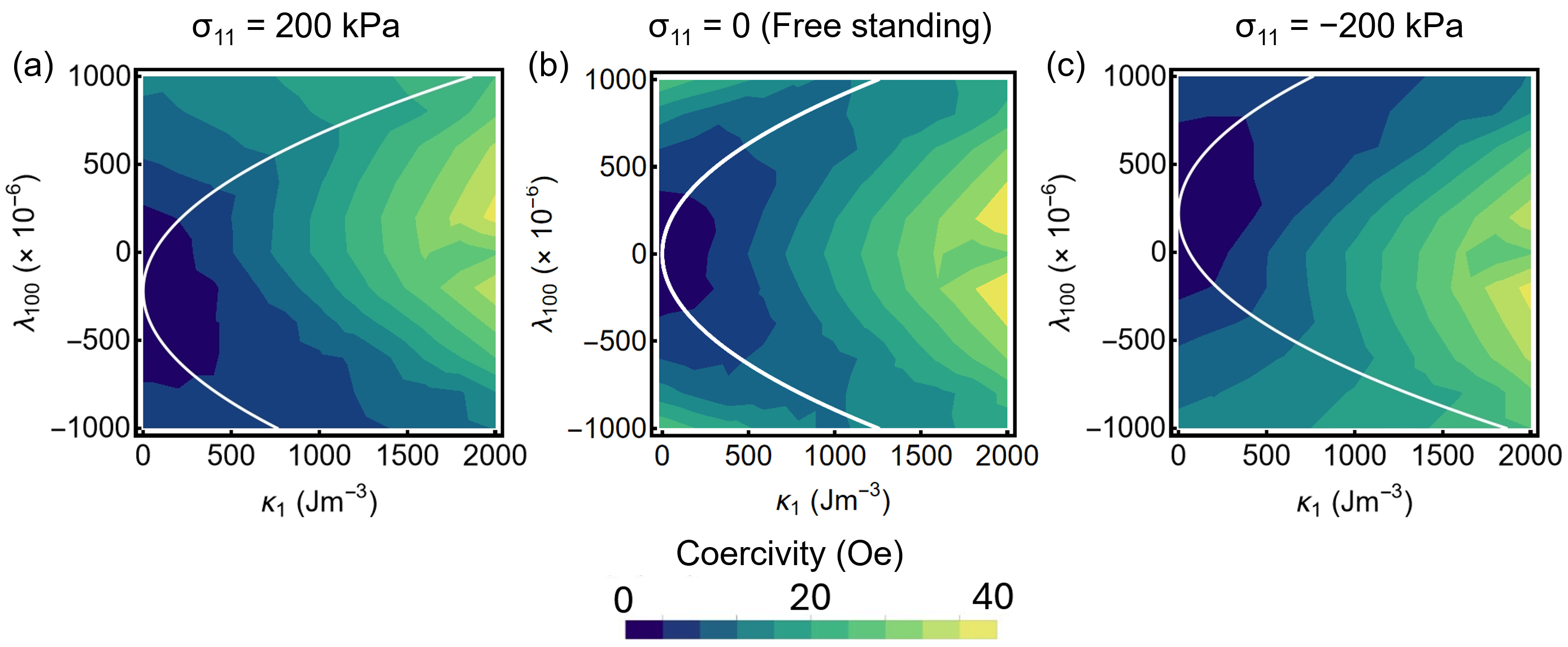}}
\caption{\label{ResidualStresses}Heatmaps showing the coercivity of magnetic alloys as a function of magnetocrystalline anisotropy and magnetostriction constants under an external stress of (a) $\sigma_{11} = 200$~$\mathrm{kPa}$, (b) $\sigma_{11} = 0$ (the free standing state for comparison) and, (c) $\sigma_{11} = -200$~$\mathrm{kPa}$. The parabolas in white represent the locus of minimum coercivity values as a function of magnetocrystalline anisotropy, magnetostriction, and residual stresses. We analytically derive these parabolic relations in section \ref{IV} and attribute their asymmetry about the $\lambda_{100}=0$ axis to the residual strains.}
\par\end{centering}
\end{figure}

Using the coercivity map with $c_{11}-c_{12} = 150$ $\mathrm{GPa}$ as a representative example Fig.~\ref{CoercivityMaps}(a), we next analyze the effect of residual stresses $\sigma_{11} = \pm200$ $\mathrm{kPa}$ on the parabolic locus for minimum coercivity, see Fig.~\ref{ResidualStresses}(a,c). We model residual stresses as an unaxial mechanical load, introduced as an additional energy term $-\sigma_{11}\varepsilon_{11}$ in Eq.~\ref{MicromagneticsEnergy}. In the absence of residual stresses, see Fig.~\ref{CoercivityMaps}(a), magnetic alloys with material constants satisfying the condition $\kappa_1 = \alpha(c_{11}-c_{12})\lambda_{100}^2$ (with $\alpha$ as a scalar constant) have minimum coercivity. This parabola is symmetric about the $\lambda_{100} = 0$ axis.
However, when small residual stresses $\sigma_{\mathrm{R}} = \sigma_{11}\hat{\mathbf{e}}_1\otimes\hat{\mathbf{e}}_1$ with $\sigma_{\mathrm{11}} = \pm$ $200$ $\mathrm{kPa}$, are applied the parabola for minimum coercivity becomes asymmetric about the $\lambda_{100} = 0$ axis, see Figs.~\ref{ResidualStresses}(a,c).  We attribute this asymmetry, or a linear shift of the parabolic locus, to the additional energy term $-\sigma_{\mathrm{R}}\cdot\vb{E} = -\sigma_{11}\varepsilon_{11}$. We predict that the small residual stresses induce a linear shifting of the minimum coercivity parabola, now expressed as $\kappa_1 = \alpha(c_{11}-c_{12})(\lambda_{100}+\beta\sigma_{11})^2$, where $\beta$ is another constant. We analytically explain this linear shift of the parabolic locus in the next section.

Overall, Figs.~\ref{CoercivityMaps}-\ref{ResidualStresses} establish coercivity maps that identify specific combinations of stiffness constants $c_{11},c_{12}$, magnetostriction $\lambda_{100}$, residual stresses $\sigma_{\mathrm{R}}$, and anisotropy constants $\kappa_1$, for which coercivity can be minimized. These maps would serve as quantitative design guidelines to design soft magnetic alloys with small hysteresis.

\section{Energy Barrier Analysis for Minimum Coercivity}\label{IV}
In this section, we theoretically analyze the effect of a localized disturbance (in the form a needle domain) and residual stresses on magnetic hysteresis. Specifically, we evaluate the energy barriers limiting the growth of a needle domain on an otherwise uniformly magnetized ellipsoid, and correlate these barriers to the width of the hysteresis loop (or coercivity). By minimizing the energy barriers governing magnetization reversal, we arrive at a mathematical relationship between material constants and applied stresses for which magnetic hysteresis is minimized. Our calculation is by no means a rigorous theoretical proof, but an analytical interpretation of the relationship between material constants and mechanical loads that collectively lower magnetic hysteresis. We support our analysis using our numerical micromagnetic computations.

%\textcolor{blue}{In this section, we will provide a physical interpretation of the observed results, focusing on the relationship between energy barriers and microstructures. These microstructures arise due to the presence of non-magnetic defects in a uniformly magnetized ellipsoid. We eventually arrive at a mathematical relation among material constants to identify conditions under which magnetic hysteresis is minimized.}

As a first step, we analyze the stability of uniform magnetization on an ellipsoid under an external field. In the following sections, we build on this analysis by introducing a needle domain and applying residual stresses to the uniformly magnetized ellipsoid. We then analyze the combined effects of these localized disturbances and mechanical loads on energy barriers governing magnetization reversal.

\subsubsection*{Linear Stability Analysis of a Uniformly Magnetized Ellipsoid}\label{LSA}
\begin{figure}
\begin{centering}
\textcolor{black}{\includegraphics[width=\textwidth]{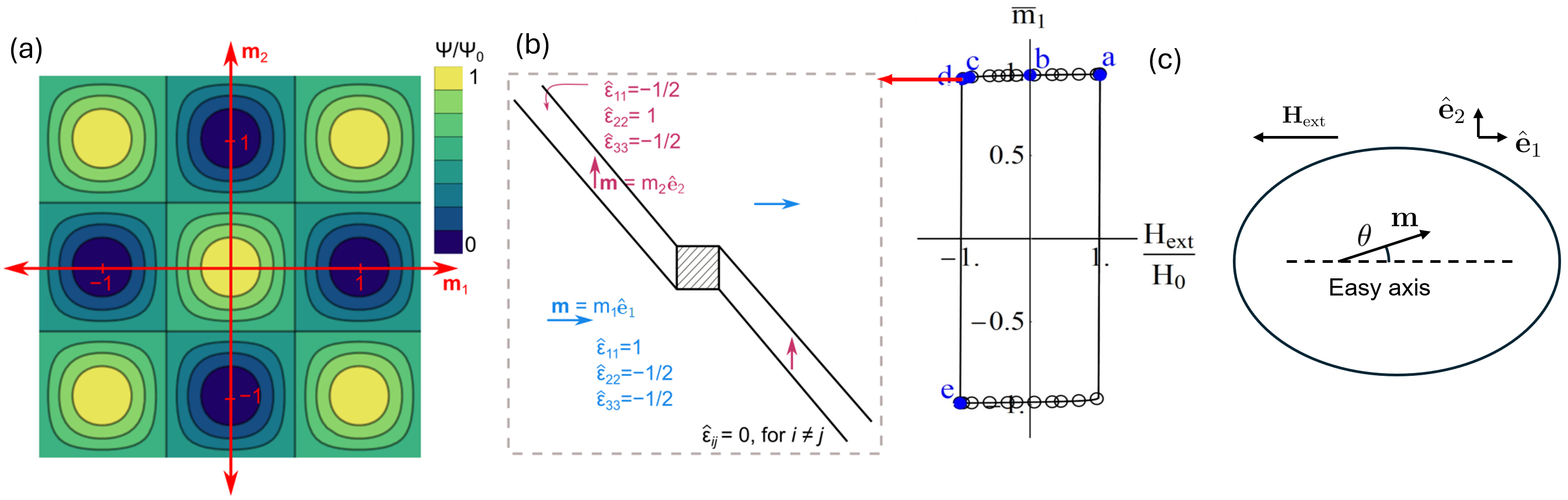}}
\caption{\label{TransientState}(a) Schematic illustration of the energy wells located at $\mathbf{m}_1 = \pm1$ and at $\mathbf{m}_2 = \pm1$. At these energy wells, the strains are minimized and correspond to the spontaneous strain values $\mathbf{E}_0 (\mathbf{m})$. (b) At the shoulder of the hysteresis loop (preceding magnetization reversal), a transient microstructural state is observed. In this state, the needle domain is extended to its maximum length and the domain walls are nearly parallel to one another. At this stage, the microstructure resembles a stripe-like domain with near-divergence-free magnetization. The ideal values of magnetization and strains in these domains correspond to the values at the energy wells, i.e., $\mathbf{m}=\mathrm{m}_1\hat{\mathbf{e}}_1$ and $\hat{\varepsilon}_{11} = 1, \hat{\varepsilon}_{22} = \hat{\varepsilon}_{33} = -\frac{1}{2}$, and $\mathbf{m}=\mathrm{m}_2\hat{\mathbf{e}}_2$ and $\hat{\varepsilon}_{22} = 1, \hat{\varepsilon}_{11} = \hat{\varepsilon}_{33} = -\frac{1}{2}$, respectively. For $i \neq j$ the shear strains are zero in both domains. (c) Illustration of a uniformly magnetized ellipsoid in linear stability analysis.}
\par\end{centering}
\end{figure}

Let us consider a large oblate ellipsoid $\mathcal{E}$ in three-dimensional space $\mathbb{R}^3$ that is uniformly magnetized along the easy axis $\mathbf{m} = \mathrm{m}_1\hat{\mathbf{e}}_1$, see Fig.~\ref{TransientState}(c). In the absence of defects, the uniform magnetization minimizes the anisotropy energy (magnetization along easy axis), the demagnetization energy (with poles far apart on the ellipsoid), and the exchange energy (no domain walls). At mechanical equilibrium, the strains on the ellipsoid $\vb{E}$ are equal to the spontaneous values $\vb{E}_0(\mathbf{m})$, and thus minimize the total elastic energy of the system. Under an external magnetic field $\mathbf{H}_{\mathrm{ext}} = -\mathrm{H_{ext}}\hat{\mathbf{e}}_1$, there is a finite Zeeman energy contribution in the system. This Zeeman energy is minimized when the magnetization on the ellipsoid reorients to align with the applied field. This reorientation, however, costs energy and perturbs the system away from equilibrium.

To analyze this increase in the free energy, we model a small and uniform in-plane perturbation $\theta \to 0$ to rotate the magnetization away from its easy axis. That is, the magnetization is:
\begin{align}
     \mathbf{m} & =\mathrm{cos}\theta~\hat{\mathbf{e}}_1 + \mathrm{sin}\theta~\hat{\mathbf{e}}_2  \nonumber\\
    & \approx \qty(1 - \dfrac{\theta^2}{2})\hat{\mathbf{e}}_1 + \theta~\hat{\mathbf{e}}_2 \quad \mathrm{for}~\theta\ll 1.
    \label{magnetizationapprox}
\end{align}
This perturbation contributes to an increase in the magnetocrystalline anisotropy energy, however, the uniform perturbation does not introduce magnetization gradients (no exchange energy). At mechanical equilibrium the elastic energy is negligible (because $\mathbf{E} = \mathbf{E}_0(\vb{m})$). Recall that we consider a large plate-like ellipsoid with a surface normal along the the $\hat{\vb{e}}_3$ direction, see Fig.~\ref{TransientState}(c). For this ellipsoid geometry the demagnetization factor is $\vb{N}_\mathrm{d} = \hat{\vb{e}}_3\otimes \hat{\vb{e}}_3$, and we neglect the contributions from the in-plane perturbations of magnetization to the demagnetization energy. Therefore, the total energy of this perturbed ellipsoid under an external field is:
\begin{align}
{\Psi}_{\theta} &= \int_{\mathcal{E}}[\kappa_1(\mathrm{m}_1^2\mathrm{m}_2^2 + \mathrm{m_2}^2\mathrm{m_3}^2 + \mathrm{m_3^2m_1^2}) - \mu_0m_s\mathbf{H}_{\mathrm{ext}}\cdot\mathbf{m}]\mathrm{d}\mathbf{x} \\& =\int_{\mathcal{E}}[\kappa_1\mathrm{cos}^2\theta\mathrm{sin}^2\theta - \mu_0m_s \mathrm{H_{ext}cos}\theta] \mathrm{d}\mathbf{x}\\
& \approx \int_{\mathcal{E}}[\kappa_1\theta^2 - \mu_0m_s \mathrm{H_{ext}}(1-\frac{\theta^2}{2})] \mathrm{d}\mathbf{x} = [\kappa_1\theta^2 - \mu_0m_s \mathrm{H_{ext}}(1-\frac{\theta^2}{2})]
\label{Psitheta}
\end{align}
The energy in Eq.~\ref{Psitheta} to the second order $\mathcal{O}(\theta^2)$ corresponds to the energy barrier limiting magnetization reversal on an ellipsoid of unit volume. At the coercivity point, the equilibrium and stability conditions are, respectively, given by:
\begin{align}
    \frac{\delta\Psi_\theta}{\delta\theta}=0, \quad \frac{\delta^2\Psi_\theta}{\delta\theta^2}\leq0.
\end{align}
The equilibrium condition is satisfied for:
\begin{align}
    \frac{\delta\Psi_\theta}{\delta\theta} &= 2\kappa_1\theta + \mu_0m_s\mathrm{H}_{\mathrm{ext}}\theta = 0,
\end{align}
and the stability criterion should be negative for magnetization reversal:
\begin{align}
    \frac{\delta^2\Psi_\theta}{\delta\theta^2} &= 2\kappa_1 + \mu_0m_s\mathrm{H}_{\mathrm{ext}} \leq 0.
\end{align}
Therefore the linear approximation of the micromagnetics energy for a uniformly magnetized domain shows the coercivity to be consistent with the results of the Stoner-Wohlfarth model \cite{doi:10.1098/rsta.1948.0007}:
\begin{align}
    |\mathrm{H_c}| & = |\mathrm{H_{ext}}| = \frac{2\kappa_1}{\mu_0m_s}.
    \label{linear approx}
\end{align}
In our analysis, we considered a second variation of the micromagnetics energy (i.e., linear stability analysis) and derived a linear relation between the coercivity and the magnetocrystalline anisotropy constant, see Eq.~\ref{linear approx}. In this analysis, we assumed a smooth ellipsoid body and a coherent rotation of the uniformly magnetized domain. A well-known difficulty of these assumptions (and therefore the linear approximation result) is that when we substitute experimentally measured values of the anisotropy constant in Eq.~\ref{linear approx} (e.g., for iron, $\kappa_1 = 4.3\times 10^3\mathrm{~J/m^3}$) the result overestimates coercivity by 3 orders of magnitude \cite{brown1963micromagnetics}. It is interesting to note that analyzing the stability of a magnetized domain, with a microstructural picture of magnetization rotation, demonstrates the dominant role of anisotropy constant on coercivity. The subtle, but important, effect of magnetostriction on magnetic hysteresis is not captured in this approach.

\subsubsection*{A Localized Disturbance on a Uniformly Magnetized Ellipsoid}

We next consider a localized disturbance (e.g., a non-magnetic inclusion defect $\Omega_d \ll \mathcal{E}$ positioned at the center of a uniformly magnetized ellipsoid and analyze the energy terms governing magnetization reversal.\footnote{Please note that this is a simplified construction of the localized disturbance when compared to the periodic array of defects used in our micormagnetic calculations. We solve the governing equations in the micromagnetics model using the Fast Fourier Transform algorithm with periodic boundary conditions in Sec.~\ref{sec:Results}.} We note that by introducing a defect, we locally perturb the uniform magnetization on the ellipsoid, and a strong demagnetization field is generated around the defect. Needle domains, as predicted by N\'eel \cite{Kurti1988} and widely observed in experiments \cite{hubert1998magnetic}, form around the defect to minimize the demagnetization energy, see Fig.~\ref{Introduction}(b). These needle domains serve as a large-localized disturbance to the uniform magnetization and grow under an external field, see Fig.~\ref{TransientState}(b). At the coercivity point, the needles have extended to their maximum length and resemble a stripe-like pattern. This microstructural picture of a growing needle domain under an external magnetic field differs from the coherent magnetization rotation considered in Sec.~\ref{LSA}. We use the transient stripe-domain microstructural state, on the shoulder of the hysteresis loop, to analyze the energy barriers limiting magnetization reversal.

In an ideal scenario, such as the infinitely long stripe-domain pattern in Fig.~\ref{TransientState}(b), the magnetization in the needle domains is uniform with $\mathbf{m}=\mathrm{m}_2\hat{\mathbf{e}}_2$ and the strains correspond to the bottom of energy wells, i.e., $\mathbf{E} = \mathbf{E}_0$. The strains inside $\hat{\mathbf{E}}_{\mathrm{in}}$ and outside $\hat{\mathbf{E}}_{\mathrm{out}}$ the needle domain are, respectively: 

\begin{equation}
\hat{\vb{E}}_{\text{in}} = \mqty(\dmat{-1/2, 1, -1/2}), \quad \hat{\vb{E}}_{\text{out}} = \mqty(\dmat{1, -1/2, -1/2}),
\label{eq:needlestrains}
\end{equation}
as shown in Fig.~\ref{TransientState}(b). For later use, we normalize the strain tensor as $\hat{\vb{E}} = \vb{E}/\lambda_{100}$. These strains for the idealized stripe-domain in Eq.~\ref{eq:needlestrains} would satisfy the kinematic compatibility condition $\hat{\mathbf{E}}_{\mathrm{in}} - \hat{\mathbf{E}}_{\mathrm{out}} = \frac{1}{2}(\mathbf{a}\otimes\hat{\mathbf{n}} + \hat{\mathbf{n}}\otimes\mathbf{a})$ for vectors $\mathbf{a}, \hat{\mathbf{n}}$. For the strains in Eq.~\ref{eq:needlestrains}, the vector $\hat{\mathbf{n}} = \langle 110\rangle$ corresponds to the domain wall orientation in the reference configuration. For these strains $\mathbf{E} = \mathbf{E}_0$ and the elastic energy in the system is zero. 

\begin{figure}
\begin{centering}
\textcolor{black}{\includegraphics[width=0.95\textwidth]{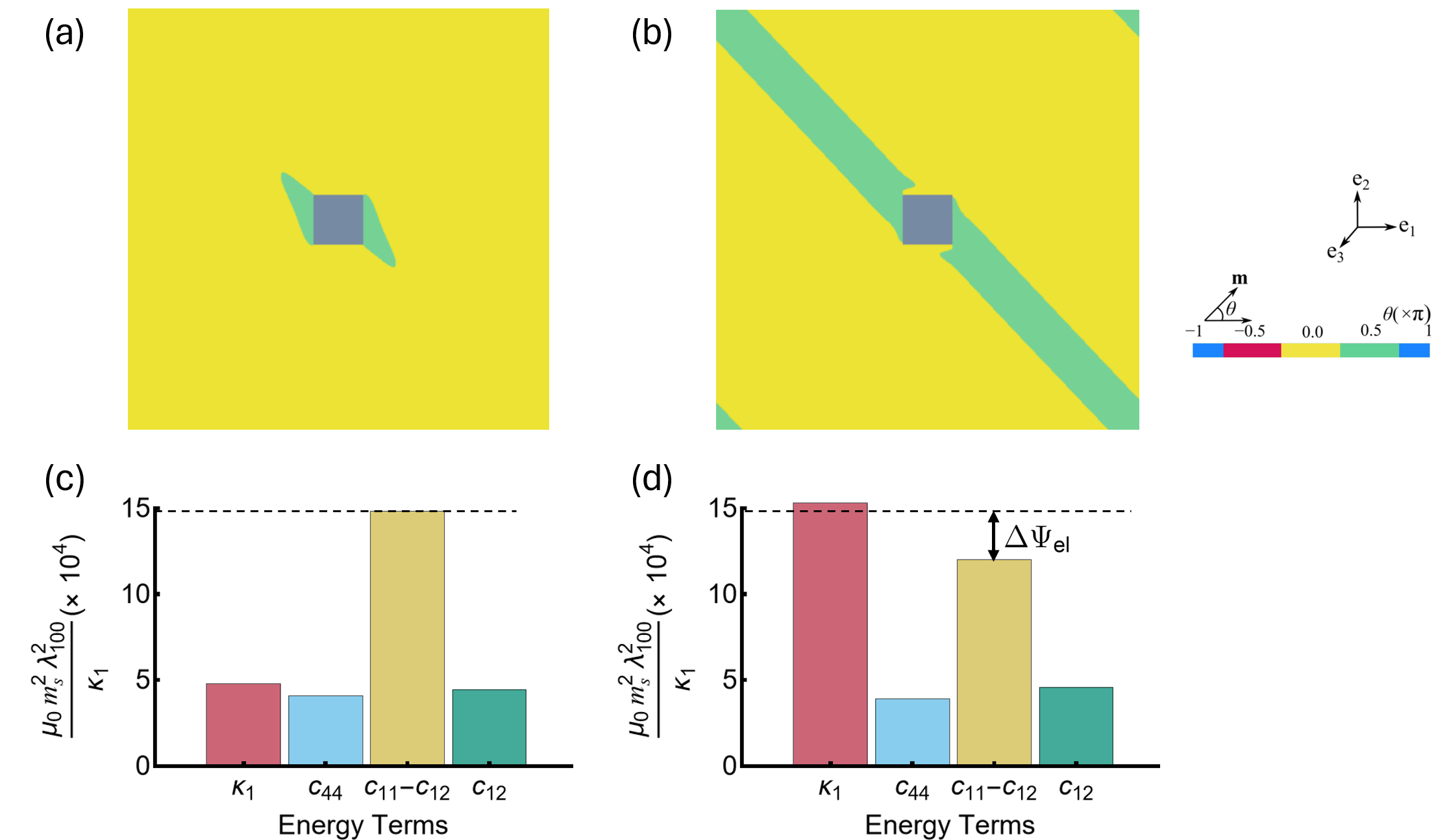}}
\caption{\label{Fig_EnergyBarrier} (a,b) The needle and stripe domain structures from an example simulation of the magnetization reversal process. The material parameter is set to be $\kappa_1 = 1000 \text{ }\mathrm{J/m^3}$, $\lambda_{100} = 800 \text{ } \mu\epsilon$ and $c_{11}-c_{12} = 150 \text{ GPa} $. The plots, from left to right, correspond to points c, and d in the coercivity loop in Fig.~\ref{Introduction}(d). (c,d) Energy contributions for configurations (a) and (b) respectively. The energies in these subfigures are evaluated in unit of $(\mu_0 m_s^2\lambda_{100}^2/\kappa_1)\times 10^{4}$. The dashed lines indicate the decrease of the elastic energy term proportional to $c_{11}-c_{12}$, $\Delta\Psi_{\mathrm{el}}$.}
\par\end{centering}
\end{figure}

In our computations, however, we note that the strains across the needle domains deviate from the spontaneous values ($\hat{\mathbf{E}} \neq \hat{\mathbf{E}}_0(\mathbf{m})$) and accumulate finite elastic energy in the system, see Fig.~\ref{Fig_EnergyBarrier}(c-d). We attribute this deviation in strains to (a) the non-zero misfit strains between the defect (a non-magnetic inclusion) and the magnetized domains, and (b) the domain walls that do not exactly satisfy the compatibility condition. These deviations in strains $\hat{\mathbf{E}}_\mathrm{D} = [\vb{E}-\vb{E}_0(\vb{m})]/\lambda_{100}$ (or, equivalently the elastic energy) contributes to the energy barrier limiting magnetization reversal:
\begin{equation}
    \int_{\mathcal{E}}\dfrac{1}{2}[\mathbf{E-E}_{0}(\mathbf{m})]\cdot\mathbb{C}[\mathbf{E-E}_{0}(\mathbf{m})] \dd \vb{x} = \int_{\mathcal{E}}\dfrac{1}{2}\lambda_{100}^2 \hat{\vb{E}}_\mathrm{D}\cdot \mathbb{C} \hat{\vb{E}}_\mathrm{D}\dd \vb{x}.\label{ElDefect}
\end{equation}
\noindent Fig.~\ref{Fig_EnergyBarrier} shows the individual contributions of the magneto-elastic and anisotropy energy terms at the needle domain (point `c' in Fig.~\ref{Introduction}(d)) and the stripe domain states (point `d' in Fig.~\ref{Introduction}(d)), respectively. In these microstructural states, the anisotropy energy is concentrated at the domain walls, and increases during magnetization reversal. However, in the presence of a defect, the magnetoelastic energy terms play an important role in lowering the energy barrier governing magnetization reversal. For example, the energy term proportional to $c_{11}-c_{12}$ decreases with the growing needle domain. This is a consequence of the lowering of lattice misfit strains along the domain walls as the longer needles/stripe domains satisfy the kinematic compatibility condition. The other elastic energy terms remain largely unchanged during magnetization reversal. Therefore, the reduction in energy contribution arising from the $c_{11}-c_{12}$ term facilitates magnetization switching and lowers coercivity. This interplay between magnetocrystalline anisotropy and magnetoelastic energies that collectively lowers coercivity was not captured in the linear stability analysis of a uniformly magnetized domain undergoing coherent magnetization rotation.\footnote{It is important to note that the coercivity $H_{\mathrm{c}}$ does not monotonously decrease as the elastic energy (scaled by $(c_{11}-c_{12})$ or $\lambda_{100}^2$) increases. A possible explanation is that when $\lambda_{100}$ is increased beyond a critical point, our previous picture of the needle domain is energetically unfavorable because of the large lattice misfit between the defect and the magnetized domain, and new intermediate states with average magnetization along ($\sqrt{\frac{2}{3}},\sqrt{\frac{1}{3}},0$) and ($\sqrt{\frac{1}{3}},\sqrt{\frac{2}{3}},0$) directions are stabilized. These intermediate states do not evolve via energy-minimizing deformations (i.e., domain structures with more compatible interfaces during evolution), and the argument of reducing the elastic energy barrier during magnetization reversal is no longer valid.}%\textcolor{blue}{This part is already included in my new analysis. This is the reason I wanted to discuss before any significant changes...}}

At the needle-domain and stripe-domain states, we note that the demagnetization energy and exchange energy are the only two energy terms containing $\nabla\mathbf{m}$. As illustrated in our previous work \cite{renuka2022design}, we find that these energy terms are only significant within the domain walls of width $l_w\sim \sqrt{A/\kappa_1}$, which is much smaller than the characteristic length of the system. For the majority of the spaces inside the domain, the divergence of magnetization is negligible. As a result, these energy terms play a minor role in governing magnetization reversal at the transient microstructural (i.e., stripe-domain) state. 
%Therefore, the demagnetization field energy is also negligible as it is minimized by long needle domains. 

%effect of other terms on energy barrier, we plot the anisotropy energy and the three magneto-elastic energy terms for the needle domain state (point c in Fig.~\ref{Introduction}(d)) and the stripe domain state (point d in Fig.~\ref{Introduction}(d)), as shown in Fig.~\ref{Fig_EnergyBarrier}. Similar to the linear stability analysis, the anisotropic energy is also increasing during magnetization reversal. In addition, the elastic energy proportional to $c_{11}-c_{12}$ also plays a role, while the other two elastic energy terms ($c_{12}$ and $c_{44}$) barely change. As a result, the elastic energy dominated by the $c_{11}-c_{12}$ term is decreasing during magnetization reversal. This could be explained as the kinematic compatibility condition being satisfied when the needle domain grows to a stripe domain. This additional energy decrease, which is scaled by $(c_{11}-c_{12})\lambda_{100}^2/(2\kappa_1)$, \textcolor{red}{lowers the energy barrier} caused by the increase of anisotropic energy. In consequence, a smaller external field will be sufficient to drive magnetization reversal, thus the coercivity will be lowered.
  
On the shoulder of the hysteresis loop, the primary energy contributions limiting the growth of the needle domain are the anisotropy and the magnetoelastic energy terms in Eqs.~\ref{Aniso},\ref{ElThreeTerms}. Let $\Delta$ denote the changes in the anisotropy and magnetoelastic energy contributions arising from an infinitesimal change of the average magnetization on the domain. These energy terms contribute to a finite energy barrier limiting magnetization reversal. We have:

\begin{equation}
    \begin{aligned}
        \Delta\Psi_{\mathrm{tot}} &= \Delta\qty(\int_{\mathcal{E}}\sum_{\substack{i,j=1 \\ i>j}}^3\mathrm{m}_i^2\mathrm{m}_j^2\dd\vb{x})+\frac{(c_{11}-c_{12})\lambda_{100}^2}{2\kappa_1}\Delta\qty(\int_{\mathcal{E}}\sum_{i=1}^3 \hat{\varepsilon}_{\mathrm{D},ii}^2\dd\vb{x}) \\
        &= \Delta F[\vb{m}]+\frac{(c_{11}-c_{12})\lambda_{100}^2}{2\kappa_1}\Delta G[\hat{\vb{E}}] \geq -\Delta \Psi_{\text{Zeeman}}= -\dfrac{\mu_0m_sH_{\mathrm{ext}}}{\kappa_1}\Delta Z[\vb{m}],
    \end{aligned} \label{StTrBarrier}
\end{equation}
in which $F[\vb{m}]$ and $Z[\vb{m}]=\int_{\mathcal{E}}\vb{m}\cdot\hat{\vb{e}}_1\dd\vb{x}$ are functionals of $\vb{m}$ and $G[\hat{\vb{E}}]$ is a functional of $\hat{\vb{E}}$. 
%$\vb{m}^{\mathrm{N}}$ and $\hat{\vb{E}}^{\mathrm{N}}$ are magnetic field and strain field at the needle domain state. 
As the needle domain extends, the anisotropy energy of the system increases ($\Delta F[\vb{m}]>0$), however, at the same time, the extended needles approach a stripe-domain pattern that approximately satisfy the kinematic compatibility condition. This compatibility reduces the total magnetoelastic energy contribution, i.e., $\Delta G[\vb{m}]<0$, see Fig.~\ref{Fig_EnergyBarrier}(c-d). This interplay between the anisotropy and magnetoelastic energy terms govern the height of the barrier, and the critical Zeeman energy necessary for magnetization reversal is:
\begin{equation}
    \Delta F[\vb{m}]+\frac{(c_{11}-c_{12})\lambda_{100}^2}{2\kappa_1}\Delta G[\hat{\vb{E}}] = -\dfrac{\mu_0m_sH_{\mathrm{c}}}{\kappa_1}\Delta Z[\vb{m}]. \label{Hc-N}
\end{equation}
This energy barrier for magnetization reversal is minimized for: 
\begin{equation}
    \dfrac{\kappa_1}{(c_{11}-c_{12})\lambda_{100}^2} = -\dfrac{\Delta G[\hat{\vb{E}}]}{2\Delta F[\vb{m}]} = \alpha. \label{alpha_eval}
\end{equation}
Eq.~\ref{alpha_eval} shows a parabolic relation between anisotropy and magnetostriction constants for which coercivity is reduced.
%\footnote{\textcolor{red}{In the supplement, we interpret the physical meaning of the We have some intuitive explanations in the supplementary information.}} 
In our numerical calculations, we find that the ratio $\alpha$ to depend on the choice of other material constants (e.g., magnetostriction $\lambda_{111}$, elastic stiffness $\mathbb{C}$) and to some extent on the defect geometry. For example, this ratio affects the width of the parabolas for minimum coercivity in Figs.~\ref{CoercivityMaps}(a-d), however, the relation $\kappa_1 \propto \lambda_{100}^2$ is a common feature in the coercivity heatmaps. %\textcolor{red}{We could also see from the above analysis that decreasing the minimum coercivity requires us to destabilize the intermediate stripe domain state.}

Overall, we compute a nonlinear stability analysis of magnetization reversal and reveal the unexpected role of magnetoelastic interactions in lowering hysteresis. The results demonstrate a parabolic relationship between anisotropy and magnetostriction constants for minimum coercivity. This subtle, but important, effect of magnetoelastic interactions on magnetic hysteresis was not captured in the prior linear stability analysis approach, and in the past has provided insights into the permalloy problem \cite{ARB2021Permalloy} and the coercivity paradox \cite{balakrishna2021tool}. We build on this result by analyzing the effect of applied stresses on magnetic hysteresis and the parabolic relation for minimum coercivity. 
%The critical point is for these two energies to be equal, yielding the observed constant $\alpha$ to be evaluated as:\footnote{If the Zeeman energy is not negligible, the value of $\alpha$ would be different. However, it is still a constant as long as the coercivity does not significantly change along the $(c_{11}-c_{12})\lambda_{100}^2/(2\kappa_1) = $ Const. parabola.}
%\begin{equation}
%    \alpha = \dfrac{\kappa_1}{(c_{11}-c_{12})\lambda_{100}^2} = -\dfrac{\Delta G[\hat{\vb{E}}]}{2\Delta F[\vb{m}]}. \label{alpha_eval}
%\end{equation}

%\textcolor{blue}{If the demagnetization is not negligible in the new phase, the conclusion $\kappa_1 = \alpha (c_{11}-c_{12})\lambda_{100}^2$ will be different. To see its effect, the demagnetization energy (normalized by $\kappa_1$) is expressed as \cite{wang2001gauss,ARB2021Permalloy}
%\begin{equation}
%\begin{aligned}
%        \Psi_{\mathrm{d}} &= \dfrac{\mu_0}{2\kappa_1}\int_{\mathbb{R}^3}|\vb{H_d}|^2\dd\vb{x} = -\dfrac{\mu_0 m_s}{2\kappa_1}\int_{\mathcal{E}}\vb{H_d}\cdot\vb{m}\dd\vb{x}, \\
%        \vb{H_d} &= -\dfrac{\mu_0 m_s}{8\pi}\grad\int_{\mathcal{E}}\grad \dfrac{1}{|\vb{x-y}|}\cdot \vb{m(y)}\dd \vb{y}.
%\end{aligned}
%\end{equation}
%Therefore, }

%\newpage

\subsection*{Effect of Residual Stress on the Energy Barrier}\label{EB-stress}
We next analyze the effect of the applied or residual stresses on the energy barrier governing the growth of the needle domain. The total elastic energy stored in the ellipsoid under a constant applied or residual stress $\sigma_\mathrm{R}$ is:
\begin{equation}
\begin{aligned}
    \psi_{\mathrm{elas}}^{\mathrm{R}} &= \int_{\mathcal{E}}\qty\Big{\dfrac{1}{2}[\mathbf{E}-\mathbf{E_{0}\mathrm{(\mathbf{m})}}]\cdot\mathbb{C}[\mathbf{E}-\mathbf{E_{0}\mathrm{(\mathbf{m})}}]-\sigma_\mathrm{R}\cdot\vb{E}}\dd\vb{x}.
\end{aligned}
\end{equation}
Using $\mathbf{E}_\mathrm{D} = \mathbf{E}-\mathbf{E}_0\mathrm{(\mathbf{m})}$, and $\mathbb{S}$ as the compliance tensor, we rewrite the elastic energy as:
\begin{equation}
\begin{aligned}
    \psi_{\mathrm{elas}}^{\mathrm{R}} &= \int_{\mathcal{E}}\qty\Big{\dfrac{1}{2}\vb{E}_{\mathrm{D}}\cdot\mathbb{C}\vb{E}_{\mathrm{D}}-\sigma_\mathrm{R}\cdot[\vb{E}_{\mathrm{D}}+\vb{E}_0(\vb{m})]}\dd\vb{x} \\
    &= \int_{\mathcal{E}}\qty\Big{\dfrac{1}{2}\vb{E}_{\mathrm{D}}\cdot\mathbb{C}\vb{E}_{\mathrm{D}}+\dfrac{1}{2}\sigma_\mathrm{R}\cdot\mathbb{S}\sigma_\mathrm{R}-\dfrac{1}{2}\sigma_\mathrm{R}\cdot\mathbb{S}\sigma_\mathrm{R}-\sigma_\mathrm{R}\cdot[\vb{E}_{\mathrm{D}}+\vb{E}_0(\vb{m})]}\dd\vb{x} \\
    &= \int_{\mathcal{E}}\qty[\dfrac{1}{2}(\vb{E}_{\mathrm{D}}-\mathbb{S}\sigma_\mathrm{R})\cdot\mathbb{C}(\vb{E}_{\mathrm{D}}-\mathbb{S}\sigma_\mathrm{R}) - \sigma_\mathrm{R}\cdot\vb{E}_0(\vb{m}) - \dfrac{1}{2}\sigma_\mathrm{R}\cdot\mathbb{S}\sigma_\mathrm{R}]\dd \vb{x} \\
    &= \int_{\mathcal{E}}\qty[\dfrac{1}{2}\vb{E}_{\mathrm{eff}}\cdot\mathbb{C}\vb{E}_{\mathrm{eff}} - \sigma_\mathrm{R}\cdot\vb{E}_0(\vb{m})]\dd \vb{x} + \mathrm{Const.}  \\
\end{aligned} \label{psi_R}
\end{equation}
In Eq.~\ref{psi_R}, the energy term $\int_\mathcal{E}-\frac{1}{2}\sigma_{\mathrm{R}}\cdot\mathbb{S}\sigma_{\mathrm{R}}\mathrm{d}\mathbf{x}$ is a constant, and we define $\vb{E}_{\mathrm{eff}} = \vb{E}_{\mathrm{D}} - \mathbb{S}\sigma_\mathrm{R}$ to be an effective strain tensor under the applied or residual stress (see section~\ref{Strain tensor-supplement} of the Supplement for strain decomposition). This effective strain accounts for heterogeneous lattice misfit strains at defects and domain walls, and homogeneous lattice distortions arising from applied mechanical stresses. Eq.~\ref{psi_R} can be viewed as the sum of an effective internal energy of the system and an external field energy $\Delta\psi = \psi_{\mathrm{R}} = -\sigma_\mathrm{R}\cdot\vb{E}_0(\vb{m})$. These applied or residual stresses affects the energy barrier governing the growth of the needle domain, and therefore, alters the coercivity of a magnet, see Fig.~\ref{ResidualStresses}(a,c). 

In our computations, we model a simple uniaxial stress on the ellipsoid $\sigma_\mathrm{R} = \sigma_{11}\hat{\mathbf{e}}_1\otimes\hat{\mathbf{e}}_1$. From Eq.~\ref{psi_R}, we note that this stress linearly scales the total elastic energy of the system (normalized by $\kappa_1$) as:

\begin{equation}
    \Psi_{\mathrm{R}} = -\dfrac{1}{\kappa_1}\int_{\mathcal{E}}\sigma_\mathrm{R}\cdot\vb{E}_0(\vb{m})\dd \vb{x} =-\dfrac{3\lambda_{100}\sigma_{11}}{2\kappa_1}\int_{\mathcal{E}}\qty(\mathrm{m}_1^2-\dfrac{1}{3})\dd \vb{x}.\label{mechanical-load-energy}
\end{equation}
The second variation of this additional energy term is positive $(\delta^2\Psi_{\mathrm{R}}/\delta\theta^2 > 0)$ if both the applied stresses and magnetostriction deform the lattices in a similar manner, that is $\sigma_{11} > 0$ and $\lambda_{100} > 0$, or, $\sigma_{11} < 0$ and $\lambda_{100} < 0$. This is because both the applied stresses and the spontaneous strains (from magnetostriction) stabilize the uniform magnetization $\vb{m}=\mathrm{m_1}\hat{\vb{e}}_1$ on the ellipsoid in an energy minimizing state. Additional energy is therefore required to perturb the system from its energy minimizing state and to grow the needle domain, which in turn corresponds to an increase in the coercivity values. By contrast, if the applied stresses and magnetostriction constants have opposing effects on lattice deformation, i.e., $\sigma_{11} > 0$ and $\lambda_{100} < 0$, or, $\sigma_{11} < 0$ and $\lambda_{100} > 0$, the coercivity of soft magnets is reduced. We use this analytical interpretation to explain the shift of the minimum coercivity parabola in Fig.~\ref{ResidualStresses} under small mechanical loads.\footnote{We consider residual stresses of the order of a few hundred kPa to be small as these loads do not significantly change the geometry of needle-domains or that of the intermediate stripe-domain patterns. We consider these domains as pre-existing nucleus in our energy barrier analysis.} 

\subsection*{Parabolic Relation with Residual Stress}

Under small external stresses of the form $\sigma_{\mathrm{R}}=\sigma_{11}\hat{\mathbf{e}}_1\otimes\hat{\mathbf{e}}_1$ with $\sigma_{11} \approx 200~\mathrm{kPa}$, the needle-domain geometry in the initial state is perturbed by a negligibly small amount.\footnote{This perturbation is negligible when compared to the change in the needle domain length with $\sigma_{11} \approx 40$~$\mathrm{MPa}$ in Fig.~\ref{CoercivityPlot}.} This small perturbation does not affect the microstructural evolution pathway during magnetization reversal. However, the applied stresses do affect the coercivity values and therefore the parabolic relation for minimum coercivity, see Fig.~\ref{ResidualStresses}. In this section, we analytically interpret the shift of the minimum coercivity parabola under applied stresses (see Figs.~\ref{ResidualStresses}(a,c)).

From Figs.~\ref{CoercivityMaps}(a-d), we note that the coercivity of a magnetic alloy $\vb{H}_{\mathrm{c}}$ is a function of the magnetocrystalline anisotropy constant $\kappa_1$ and the magnetostriction constant $\lambda_{100}$.\footnote{Please note that other material parameters such as the stiffness tensor, magnetostriction along $\{111\}$, saturation magnetization $m_s$ were held constant in our computations.} We also note that this coercivity is affected by the applied or residual stresses, see Figs.~\ref{ResidualStresses}(a,c). Based on Figs.~\ref{CoercivityMaps},~\ref{ResidualStresses}, we approximate the effective coercivity $\vb{H}_{\mathrm{c}}$, as a sum of the intrinsic coercivity $\vb{H}_{\kappa, \lambda}$, determined by material constants, and a coercivity shift $\mathbf{H}_{\sigma}$, induced by the applied mechanical stresses.

%\begin{equation}
%    |\vb{H}_{\mathrm{c}}| = |\mathbf{H}_{\kappa,\lambda}| \pm|\mathbf{H}_{\sigma}|
%    \label{effective-coercivity}
%\end{equation}

%\begin{align}
%    \delta^2\Psi_{\mathrm{Zeeman}} & \approx \delta^2\Psi_{\mathrm{R}}
%\end{align}

As discussed in Sec.~\ref{EB-stress}, the applied stresses increase or decrease the energy barrier governing magnetization reversal, see Eq.~\ref{mechanical-load-energy}. For sufficiently small mechanical loads---which neither significantly change the needle domain geometry nor the magnetization reversal pathway---we correlate the change in elastic energy arising from the applied or residual stresses to the change in the Zeeman energy necessary for magnetization reversal. 

Let us assume that the residual stresses perturb the average magnetization on the ellipsoid $\langle \vb{m} \rangle$ by an infinitesimal angle $\delta\theta \ll 1$. The corresponding changes to the Zeeman and the residual elastic energy on an ellipsoid of unit volume at $\theta\approx 0$, respectively, are:
\begin{align}
    \Delta\Psi_{\mathrm{Zeeman}} &= \dfrac{1}{2}\delta^2\Psi_{\mathrm{Zeeman}} = \dfrac{\mu_0m_s|\vb{H}_{\sigma}|}{2\kappa_1}\delta^2(\cos\theta) 
    = -\dfrac{\mu_0m_s|\vb{H}_\sigma|}{2\kappa_1}\delta\theta^2 \label{Zeeman-Stress} \\ 
    \Delta\Psi_{\mathrm{R}} &= \dfrac{1}{2}\delta^2\Psi_{\mathrm{R}} = -\dfrac{3\lambda_{100}\sigma_{11}}{4\kappa_1}\delta^2(\cos^2\theta) 
    = \dfrac{3\lambda_{100}\sigma_{11}}{2\kappa_1}\delta\theta^2. \label{Residual-Stress}
\end{align}
From Eqs.~\ref{Zeeman-Stress}-\ref{Residual-Stress}, the coercivity shift induced by applied stresses is given by $\vb{H}_{\sigma} \approx -(3\lambda_{100}\sigma_{11} / \mu_0 m_s)\hat{\vb{e}}_1$.
For a constant magnetocrystalline anisotropy and a constant value of the applied mechanical stress, we approximate the effective coercivity function as a Taylor series polynomial at $\lambda_{100} = \lambda^*$ and up to second order $\mathcal{O}[(\lambda_{100}-\lambda^*)^2]$:
\begin{equation}
    |\vb{H}_{\mathrm{c}}| \approx |\vb{H}^*| + \pdv[2]{|\vb{H}_{\mathrm{c}}|}{\lambda_{100}}\qty(\lambda_{100}-\lambda^*)^2.\label{coercivityfunction}
\end{equation}
In Eq.~\ref{coercivityfunction}, we assume the function $|\mathbf{H}_c|(\lambda_{100})$ to be twice differentiable at $\lambda_{100} = \lambda^*$, and from Fig.~\ref{CoercivityMaps}, we note that the local minimum in the coercivity maps satisfies $\partial|\vb{H}_{c}|/\partial\lambda_{100}=0$. In the presence of small residual stresses, and using $\vb{H}_{\sigma} \approx -(3\lambda_{100}\sigma_{11} / \mu_0 m_s)\hat{\vb{e}}_1$, the coercivity function can be approximated as:
%With the additional effective field, the total effect of the two fields resembles the minimum coercivity parabola without residual stress and follows the same argument as Eqs.~\ref{Hc-N}-\ref{alpha_eval}. 
%Assuming that other energy terms are unchanged, the only difference is in changing $H_{\mathrm{c}} = |\vb{H}_{\mathrm{c}}|$ to $|\vb{H}_{\mathrm{c}}+\vb{H}_\sigma|$ in these equations.
\begin{align}
    |\vb{H}_{\kappa,\lambda} + \vb{H}_{\sigma}| &\approx |\vb{H}^*| + \pdv[2]{|\vb{H}_{\mathrm{c}}|}{\lambda_{100}}\qty(\lambda_{100}-\lambda^*)^2\\ \nonumber
    |\vb{H}_{\kappa,\lambda}| & = |\vb{H}^*| + \pdv[2]{|\vb{H}_{\mathrm{c}}|}{\lambda_{100}}\qty(\lambda_{100}-\lambda^*)^2 + \frac{3\lambda_{100}\sigma_{11}}{\mu_0 m_s}\\ \nonumber
    & = k\bigg(\lambda_{100} - \lambda^* + \frac{3\lambda_{100}\sigma_{11}}{2k \mu_0 m_s}\bigg)^2 + \mathrm{Const.}~\forall~k = \pdv[2]{|\vb{H}_{\mathrm{c}}|}{\lambda_{100}} \ne 0.
    \label{parabolic-shift}
\end{align}
The intrinsic coercivity in Eq.~\ref{parabolic-shift} $|\vb{H}_{\kappa, \lambda}|$ is minimized for 
\begin{align}
    \lambda_{100}-\lambda^* + \frac{3 \sigma_{11}}{2k\mu_0 m_s} = 0, \quad \text{or} \\
    \lambda_{100} = \lambda^* -\frac{3 \sigma_{11}}{2k\mu_0 m_s}. \label{linearshift-stress}
\end{align}
In Eq.~\ref{linearshift-stress}, we view $-\frac{3 \sigma_{11}}{2k\mu_0 m_s} = -\beta\sigma_{11}$ as a linear shift of the parabolic locus for minimum coercivity (for some constants $\beta, k$).\footnote{The constant $k = \pdv[2]{|\vb{H}_{\mathrm{c}}|}{\lambda_{100}} \ne 0$ is approximately independent of $\lambda_{100}$, as observed from Fig.~\ref{ResidualStresses}, and we provide its specific form and interpret its physical meaning in the supplementary information.} That is, from Eq.~\ref{alpha_eval} and Eq.~\ref{linearshift-stress}, we have the parabolic locus for minimum coercivity to be:
\begin{equation}
    \kappa_1 = \alpha(c_{11}-c_{12})(\lambda_{100}+\beta\sigma_{11})^2, \label{LinearShift}
\end{equation}
for some constants $\alpha, \beta$ that would depend on the geometry of the inclusion defect(s) and other magnetic material parameters that are held as constants in one of our coercivity maps. The shift of the parabolic locus given by Eq.~\ref{LinearShift} is consistent with our numerical calculations in Fig.~\ref{ResidualStresses} under applied stresses $\sigma_{11} = \pm200~\mathrm{kPa}$ (see Section~\ref{Shift-parabola} of the supplement for details).

\section{Discussion}
%500 words
In this work, we investigated whether and how magnetoelastic interactions, which are commonly neglected in the study of magnetic hysteresis, affect coercivity. Our theoretical analysis shows that magnetoelastic interactions, for suitable values of magnetostriction and applied stresses, dominate magnetization reversal and can be designed to reduce magnetic hysteresis. In Study 1, we show that the coercivity of an alloy can be reduced (i.e., comparable to permalloy) despite its large anisotropy constant at suitable values of magnetostriction constant $\lambda_{100}$ and residual stresses $\sigma_{\mathrm{R}}$. In Study 2, we use a theory-guided search of the material parameter space to establish coercivity maps as a function of magnetoelastic constants ($\lambda_{100}, c_{11}, c_{12}$) and applied stresses ($\sigma_{\mathrm{R}}$). These calculations establish coercivity heatmaps that provide quantitative guidelines to design magnets with small hysteresis. Finally, we theoretically analyze the energy barriers limiting the growth of a needle domain (under applied stresses) on magnetic hysteresis. Our results identify a parabolic relation between material constants and applied stresses $\kappa_1 = \alpha(c_{11}-c_{12})(\lambda_{100}+\beta\sigma_{11})^2$ for which coercivity is minimized. Below, we discuss some limiting conditions on our results and then highlight the key features of our work.

\subsection*{Limitations}
In our micromagnetic calculations, we model a finite computational domain $\Omega$ embedded within an infinitely large ellipsoid $\mathcal{E}$. We numerically solve the governing equations Eq.~\ref{LLG}--\ref{eq:mechanicaleq} on $\Omega$ using a Fast Fourier Transform (FFT) algorithm, and analytically calculate the geometry-dependent demagnetization fields. The FFT introduces periodic boundary conditions on $\Omega$, which requires careful consideration of the computational domain to minimize stray field interactions between periodic images. To mitigate this effect, we select defect and domain sizes such that $V(\Omega_d) \ll V(\Omega)$, ensuring that the locally varying demagnetization field decays near computational domain boundaries, i.e., $\tilde{\mathbf{H}}_d \to \vb{0}$ (see \cite{balakrishna2021tool, renuka2024Rethinking} for further details). Another limitation in our study is the assumption of homogeneous and uniaxial applied stress when computating coercvity. In reality, applied or residual stresses---often localized and inhomogeneous, particularly near defects and precipitates---can accumulate during material synthesis or when integrating magnetic alloys into devices. These stress distributions, along with defect geometries and needle domain distribution, could affect our theoretical predictions of coercivity. Additionally, in our computations, we independently vary the magnetostriction constant $\lambda_{100}$ while keeping the elastic stiffness constants fixed. However, during alloy synthesis, altering the magnetostriction (e.g., via substitutional doping) may impact the elastic stiffness and vice versa. The interdependence between material constants is not fully understood, and a broader combinatorial investigation, using experimental validation, first-principles calculations, or model Hamiltonian approaches would be valuable. With these caveats in mind, we next highlight the key findings of our work.

\subsection*{Significance}
The distinguishing feature of our micromagnetics work is that we correlate magnetic coercivity with the energy barriers governing the growth of needle domains. 
%These needle domains form around a non-magnetic inclusion defect, embedded within a uniformly magnetized ellipsoid, and grow under an external field. %\textcolor{blue}{in addition to a uniformly magnetized ellipsoid, we introduce a non-magnetic defect with magnetostriction} in our numerical micromagnetics to compute coercivity. 
In so doing, we capture the nuanced interplay between magnetocrystalline anisotropy and magnetostriction constants in lowering magnetic hysteresis. We find that the magnetization reverses readily in magnetic alloys with suitable magnetostrictive coefficients, despite their large anisotropy constants. This finding contrasts with some reports in the literature, in which the small magnetostriction (in addition to small anisotropy) is considered necessary for lowering hysteresis \cite{10.1063/1.369539,PhysRevB.66.174419,HERZER200599}. While small magnetostriction and anisotropy constants contribute to small hysteresis and lowering these constants to near zero values has been the current strategy for developing soft magnetic alloys, as shown in Fig.~\ref{Introduction}(a), our results expand the material parameter space to search for new combinations of magnetic constants (including $\lambda_{100}, \mathbb{C}, \sigma_{\mathrm{R}}$) with small hysteresis. Another feature of our work is that in addition to designing fundamental material constants (e.g., $\lambda_{100}, \mathbb{C}$) to reduce hysteresis, we show that one can introduce mechanical constraints, such as compressive or tensile stresses to tailor coercivity. Furthermore, we show how applied or residual stresses affect the shape of the hysteresis loops, which would be relevant to the application of magnetic materials in engineering memory and actuator devices. Overall, we demonstrate how applied or residual stresses can be engineered to optimize the energy barriers governing magnetization reversal, and theoretically arrive a mathematical relation $\kappa_1 = \alpha(c_{11}-c_{12})(\lambda_{100}+\beta\sigma_{11})^2$ for minimum coercivity. This relation could serve as a quantitative design principle in the discovery of new soft magnetic alloys.

\section{Conclusion}
%100 words
To conclude, we present a micromagnetics framework based on magnetostriction and localized disturbances as a way forward to computing hysteresis in magnetic alloys. Using this approach, we show that magnetoelastic interactions, arising from magnetostriction, elastic stiffness constants, and applied or residual stresses, affect the shape and width of hysteresis loops. Our energy barrier analysis shows that the magnetoelastic interactions can be quantitatively designed following the $\kappa_1 = \alpha(c_{11}-c_{12})(\lambda_{100}+\beta\sigma_{11})^2$ design rule to reduce magnetic hysteresis despite large anisotropy constants.
\vspace{2mm}

\section*{Acknowledgement}
H. Guan, N. Ahani, and A. Renuka Balakrishna gratefully acknowledge research funding by the U.S. Department of Energy, Office of Basic Energy Sciences, Division of Materials Sciences and Engineering under Award DE-SC0024227 (HG), and the Air Force Fiscal Year 2023 Young Investigator Research Program, United States under Grant No. FOA-AFRL-AFOSR-2022-0005 (HG, ARB). The authors acknowledge the Center for Scientific Computing at University of California, Santa Barbara (MRSEC; NSF DMR 2308708) for providing computational resources that contributed to the results reported in this paper.

%\section*{CRediT authorship contribution statement} %Acta Materilia format
%\textbf{Hongyi Guan:} Micromagnetics calculations, Theoretical Analysis, Writing – original draft, Writing – review \& editing. \textbf{Negar Ahani:} Micromagnetics calculations, Writing – original draft \textbf{Ananya Renuka Balakrishna:}  Funding acquisition, Resources, Supervision, Writing – review \& editing.

\bibliographystyle{elsarticle-num}
\bibliography{references}

\begin{thebibliography}{10}
\expandafter\ifx\csname url\endcsname\relax
  \def\url#1{\texttt{#1}}\fi
\expandafter\ifx\csname urlprefix\endcsname\relax\def\urlprefix{URL }\fi
\expandafter\ifx\csname href\endcsname\relax
  \def\href#1#2{#2} \def\path#1{#1}\fi

\bibitem{silveyra2018soft}
J.~M. Silveyra, E.~Ferrara, D.~L. Huber, T.~C. Monson, Soft magnetic materials
  for a sustainable and electrified world, Science 362~(6413) (2018).
\newblock \href {https://doi.org/10.1126/science.aao0195}
  {\path{doi:10.1126/science.aao0195}}.

\bibitem{gutfleisch2011magnetic}
M.~A. Willard, E.~Br{\"u}ck, C.~H. Chen, S.~Sankar, J.~P. Liu, Magnetic
  materials and devices for the 21st century: stronger, lighter, and more
  energy efficient, Adv. Mater. 23~(7) (2011) 821--842.
\newblock \href {https://doi.org/10.1002/adma.201002180}
  {\path{doi:10.1002/adma.201002180}}.

\bibitem{bertotti2005science}
G.~Bertotti, I.~D. Mayergoyz, The Science of Hysteresis, Elsevier, 2005.

\bibitem{james2015magnetic}
R.~D. James, Magnetic alloys break the rules, Nature 521~(7552) (2015)
  298--299.
\newblock \href {https://doi.org/10.1038/521298a} {\path{doi:10.1038/521298a}}.

\bibitem{Foggiatto2022}
A.~L. Foggiatto, S.~Kunii, C.~Mitsumata, M.~Kotsugi, Feature extended energy
  landscape model for interpreting coercivity mechanism, Commun. Phys. 5~(1)
  (2022) 277.
\newblock \href {https://doi.org/10.1038/s42005-022-01054-3}
  {\path{doi:10.1038/s42005-022-01054-3}}.

\bibitem{jiles1986theory}
D.~C. Jiles, D.~L. Atherton, Theory of ferromagnetic hysteresis, J. Magn. Magn.
  Mater. 61~(1-2) (1986) 48--60.
\newblock \href {https://doi.org/10.1016/0304-8853(86)90066-1}
  {\path{doi:10.1016/0304-8853(86)90066-1}}.

\bibitem{kersten1943underlying}
M.~Kersten, Underlying theory of ferromagnetic hysteresis and coercivity, in:
  H. Leipzig (Ed.) (1943).

\bibitem{Neel1981}
L.~N{\'e}el, Some unresolved problems in magnetism, IEEE Trans. Magn. 17~(6)
  (1981) 2516--2519.
\newblock \href {https://doi.org/10.1109/TMAG.1981.1061752}
  {\path{doi:10.1109/TMAG.1981.1061752}}.

\bibitem{han2023strong}
L.~Han, F.~Maccari, I.~Soldatov, N.~J. Peter, I.~R. Souza~Filho,
  R.~Sch{\"a}fer, O.~Gutfleisch, Z.~Li, D.~Raabe, Strong and ductile high
  temperature soft magnets through widmanst{\"a}tten precipitates, Nat. Commun.
  14~(1) (2023) 8176.

\bibitem{Kunii2022}
S.~Kunii, K.~Masuzawa, A.~L. Fogiatto, C.~Mitsumata, M.~Kotsugi, Causal
  analysis and visualization of magnetization reversal using feature extended
  landau free energy, Sci. Rep. 12~(1) (2022) 19892.
\newblock \href {https://doi.org/10.1038/s41598-022-21971-1}
  {\path{doi:10.1038/s41598-022-21971-1}}.

\bibitem{schaffer2021machine}
S.~Schaffer, N.~J. Mauser, T.~Schrefl, D.~Suess, L.~Exl, Machine learning
  methods for the prediction of micromagnetic magnetization dynamics, IEEE
  Transactions on Magnetics 58~(2) (2021) 1--6.
\newblock \href {https://doi.org/10.1109/TMAG.2021.3095251}
  {\path{doi:10.1109/TMAG.2021.3095251}}.

\bibitem{bozorth1993ferromagnetism}
R.~M. Bozorth, Ferromagnetism, Wiley-VCH, 1993.

\bibitem{bozorth1953permalloy}
R.~M. Bozorth, The permalloy problem, Rev. Mod. Phys. 25~(1) (1953) 42.
\newblock \href {https://doi.org/10.1103/RevModPhys.25.42}
  {\path{doi:10.1103/RevModPhys.25.42}}.

\bibitem{hubert1998magnetic}
A.~Hubert, R.~Sch{\"a}fer, Magnetic Domains: The Analysis of Magnetic
  Microstructures, Springer, 1998.

\bibitem{wu2024effect}
X.~Wu, X.~Li, S.~Li, Y.~Li, Effect of \ce{Co} content on properties of
  \ce{Fe-B-Nb-Y} bulk metallic glasses, Met. Mater. Int. (2024) 1--6.~\href
  {https://doi.org/10.1007/s12540-023-01589-2}
  {\path{doi:10.1007/s12540-023-01589-2}}.

\bibitem{mchenry1999amorphous}
M.~E. McHenry, M.~A. Willard, D.~E. Laughlin, Amorphous and nanocrystalline
  materials for applications as soft magnets, Prog. Mater. Sci. 44~(4) (1999)
  291--433.
\newblock \href {https://doi.org/10.1016/S0079-6425(99)00002-X}
  {\path{doi:10.1016/S0079-6425(99)00002-X}}.

\bibitem{Mitra_2024}
A.~Mitra, J.~Mohapatra, M.~Aslam, Magnetic and electronic properties of
  anisotropic magnetite nanoparticles, Materials Research Express 11~(2) (2024)
  022002.
\newblock \href {https://doi.org/10.1088/2053-1591/ad2a84}
  {\path{doi:10.1088/2053-1591/ad2a84}}.

\bibitem{Rafique2004magnetic}
S.~Rafique, J.~R. Cullen, M.~Wuttig, J.~Cui, Magnetic anisotropy of \ce{FeGa}
  alloys, J. Appl. Phys. 95~(11) (2004) 6939--6941.
\newblock \href {https://doi.org/10.1063/1.1676054}
  {\path{doi:10.1063/1.1676054}}.

\bibitem{Clark2003}
A.~E. Clark, K.~B. Hathaway, M.~Wun-Fogle, J.~B. Restorff, T.~A. Lograsso,
  V.~M. Keppens, G.~Petculescu, R.~A. Taylor, Extraordinary magnetoelasticity
  and lattice softening in \uppercase{bcc} \ce{FeGa} alloys, J. Appl. Phys.
  93~(10) (2003) 8621--8623.
\newblock \href {https://doi.org/10.1063/1.1540130}
  {\path{doi:10.1063/1.1540130}}.

\bibitem{Clemente2017}
C.~Clemente, A.~Mahgoub, D.~Davino, C.~Visone, Multiphysics circuit of a
  magnetostrictive energy harvesting device, J. Intell. Mater. Syst. Struct. 28
  (2017) 1045389X1668544.
\newblock \href {https://doi.org/10.1177/1045389X16685444}
  {\path{doi:10.1177/1045389X16685444}}.

\bibitem{TICKLE1999627}
R.~Tickle, R.~D. James, Magnetic and magnetomechanical properties of
  \ce{Ni2MnGa}, J. Magn. Magn. Mater. 195~(3) (1999) 627--638.
\newblock \href {https://doi.org/10.1016/S0304-8853(99)00292-9}
  {\path{doi:10.1016/S0304-8853(99)00292-9}}.

\bibitem{Hu2001Magnetic}
F.-X. Hu, J.-R. Sun, G.-H. Wu, B.-G. Shen, Magnetic entropy change in
  \ce{Ni_{50.1}Mn_{20.7}Ga_{29.6}} single crystal, J. Appl. Phys. 90~(10)
  (2001) 5216--5219.
\newblock \href {https://doi.org/10.1063/1.1410890}
  {\path{doi:10.1063/1.1410890}}.

\bibitem{james1998magnetostriction}
R.~D. James, M.~Wuttig, Magnetostriction of martensite, Phil. Mag. 77~(5)
  (1998) 1273--1299.
\newblock \href {https://doi.org/10.1080/01418619808214252}
  {\path{doi:10.1080/01418619808214252}}.

\bibitem{miehe2011incremental}
C.~Miehe, B.~Kiefer, D.~Rosato, An incremental variational formulation of
  dissipative magnetostriction at the macroscopic continuum level, Int. J.
  Solids Struct. 48~(13) (2011) 1846--1866.
\newblock \href {https://doi.org/10.1016/j.ijsolstr.2011.02.011}
  {\path{doi:10.1016/j.ijsolstr.2011.02.011}}.

\bibitem{brown1966magnetoelastic}
W.~F. Brown, Magnetoelastic Interactions, Springer tracts in natural
  philosophy, Springer-Verlag, 1966.

\bibitem{tsukahara2022role}
H.~Tsukahara, H.~Imamura, C.~Mitsumata, K.~Suzuki, K.~Ono, Role of
  magnetostriction on power losses in nanocrystalline soft magnets, NPG Asia
  Mater. 14~(1) (2022) 1--12.
\newblock \href {https://doi.org/10.1038/s41427-022-00388-2}
  {\path{doi:10.1038/s41427-022-00388-2}}.

\bibitem{wang2001gauss}
X.-P. Wang, C.~J. Garc{\i}a-Cervera, W.~E, A gauss--seidel projection method
  for micromagnetics simulations, J. Comput. Phys. 171~(1) (2001) 357--372.
\newblock \href {https://doi.org/10.1006/jcph.2001.6793}
  {\path{doi:10.1006/jcph.2001.6793}}.

\bibitem{chopra2015non}
H.~D. Chopra, M.~Wuttig, Non-joulian magnetostriction, Nature 521~(7552) (2015)
  340--343.
\newblock \href {https://doi.org/10.1038/nature14459}
  {\path{doi:10.1038/nature14459}}.

\bibitem{dabade2019micromagnetics}
V.~Dabade, R.~Venkatraman, R.~D. James, Micromagnetics of galfenol, J.
  Nonlinear Sci. 29~(2) (2019) 415--460.
\newblock \href {https://doi.org/10.1007/s00332-018-9492-8}
  {\path{doi:10.1007/s00332-018-9492-8}}.

\bibitem{yang2021recent}
Y.~Yang, T.~Liu, L.~Bi, L.~Deng, Recent advances in development of magnetic
  garnet thin films for applications in spintronics and photonics, J. Alloys
  Compd. 860 (2021) 158235.
\newblock \href {https://doi.org/10.1016/j.jallcom.2020.158235}
  {\path{doi:10.1016/j.jallcom.2020.158235}}.

\bibitem{kubota2012stress}
M.~Kubota, A.~Tsukazaki, F.~Kagawa, K.~Shibuya, Y.~Tokunaga, M.~Kawasaki,
  Y.~Tokura, Stress-induced perpendicular magnetization in epitaxial iron
  garnet thin films, Appl. Phys. Express 5~(10) (2012) 103002.
\newblock \href {https://doi.org/10.1143/APEX.5.103002}
  {\path{doi:10.1143/APEX.5.103002}}.

\bibitem{ciubotariu2019strain}
O.~Ciubotariu, A.~Semisalova, K.~Lenz, M.~Albrecht, Strain-induced
  perpendicular magnetic anisotropy and gilbert damping of \ce{Tm3Fe5O12} thin
  films, Sci. Rep. 9~(1) (2019) 1--8.
\newblock \href {https://doi.org/10.1038/s41598-019-53255-6}
  {\path{doi:10.1038/s41598-019-53255-6}}.

\bibitem{lewis1964permalloy}
B.~Lewis, The permalloy problem and anisotropy in
  \uppercase{n}ickel-\uppercase{i}ron magnetic films, Br. J. Appl. Phys. 15~(5)
  (1964) 531.
\newblock \href {https://doi.org/10.1088/0508-3443/15/5/310}
  {\path{doi:10.1088/0508-3443/15/5/310}}.

\bibitem{becker1939ferromagnetismus}
R.~Becker, W.~D{\"o}ring, Ferromagnetismus (\uppercase{F}erromagnetism),
  Berlin, Germany (1939).

\bibitem{kittel1956ferromagnetic}
C.~Kittel, J.~K. Galt, Ferromagnetic domain theory, in: Solid state physics,
  Vol.~3, Elsevier, 1956, pp. 437--564.
\newblock \href {https://doi.org/10.1016/S0081-1947(08)60136-8}
  {\path{doi:10.1016/S0081-1947(08)60136-8}}.

\bibitem{schafer2020tomography}
R.~Sch{\"a}fer, S.~Schinnerling, Tomography of basic magnetic domain patterns
  in ironlike bulk material, Phys. Rev. B 101~(21) (2020) 214430.
\newblock \href {https://doi.org/10.1103/PhysRevB.101.214430}
  {\path{doi:10.1103/PhysRevB.101.214430}}.

\bibitem{PhysRevMaterials.2.014412}
Y.~He, J.~M.~D. Coey, R.~Schaefer, C.~Jiang, Determination of bulk domain
  structure and magnetization processes in bcc ferromagnetic alloys: Analysis
  of magnetostriction in
  $\mathrm{F}{\mathrm{e}}_{83}\mathrm{G}{\mathrm{a}}_{17}$, Phys. Rev. Mater. 2
  (2018) 014412.
\newblock \href {https://doi.org/10.1103/PhysRevMaterials.2.014412}
  {\path{doi:10.1103/PhysRevMaterials.2.014412}}.

\bibitem{balakrishna2021tool}
A.~R.~Balakrishna, R.~D. James, A tool to predict coercivity in magnetic
  materials, Acta Mater. 208 (2021) 116697.
\newblock \href {https://doi.org/10.1016/j.actamat.2021.116697}
  {\path{doi:10.1016/j.actamat.2021.116697}}.

\bibitem{renuka2022design}
A.~R.~Balakrishna, R.~D. James, Design of soft magnetic materials, npj Comput.
  Mater. 8~(1) (2022) 1--10.
\newblock \href {https://doi.org/10.1038/s41524-021-00682-7}
  {\path{doi:10.1038/s41524-021-00682-7}}.

\bibitem{ARB2021Permalloy}
A.~R.~Balakrishna, R.~D. James, A solution to the permalloy problem--a
  micromagnetic analysis with magnetostriction, Appl. Phys. Lett. 118~(21)
  (2021) 212404.
\newblock \href {https://doi.org/10.1063/5.0051360}
  {\path{doi:10.1063/5.0051360}}.

\bibitem{brown1963micromagnetics}
W.~F. Brown, Micromagnetics, Interscience tracts on physics and astronomy,
  Interscience Publishers, 1963.

\bibitem{PhysRevB.105.214431}
J.~Mohapatra, J.~Fischbacher, M.~Gusenbauer, M.~Y. Xing, J.~Elkins, T.~Schrefl,
  J.~P. Liu, Coercivity limits in nanoscale ferromagnets, Phys. Rev. B 105
  (2022) 214431.
\newblock \href {https://doi.org/10.1103/PhysRevB.105.214431}
  {\path{doi:10.1103/PhysRevB.105.214431}}.

\bibitem{Panduranga2021}
K.~Mohanchandra, Z.~Xiao, J.~Schneider, L.~Taehwan, C.~Klewe, R.~Chopdekar,
  P.~Shafer, A.~N'Diaye, E.~Arenholz, R.~Candler, G.~Carman, Single magnetic
  domain \uppercase{T}erfenol-\uppercase{D} microstructures with passivating
  oxide layer, J. Magn. Magn. Mater. 528 (2021) 167798.
\newblock \href {https://doi.org/10.1016/j.jmmm.2021.167798}
  {\path{doi:10.1016/j.jmmm.2021.167798}}.

\bibitem{meisenheimer2021engineering}
P.~B. Meisenheimer, R.~A. Steinhardt, S.~H. Sung, L.~D. Williams, S.~Zhuang,
  M.~E. Nowakowski, S.~Novakov, M.~M. Torunbalci, B.~Prasad, C.~J. Zollner,
  et~al., Engineering new limits to magnetostriction through metastability in
  iron-gallium alloys, Nat. Commun. 12~(1) (2021) 1--8.
\newblock \href {https://doi.org/10.1038/s41467-021-22793-x}
  {\path{doi:10.1038/s41467-021-22793-x}}.

\bibitem{chakrabarti2022magnetic}
S.~Chakrabarti, A.~Vilan, G.~Deutch, A.~Oz, O.~Hod, J.~E. Peralta, O.~Tal,
  Magnetic control over the fundamental structure of atomic wires, Nat. Commun.
  13~(1) (2022) 1--12.
\newblock \href {https://doi.org/doi.org/10.1038/s41467-022-31456-4}
  {\path{doi:doi.org/10.1038/s41467-022-31456-4}}.

\bibitem{hsieh2021strain}
C.-Y. Hsieh, P.-C. Jiang, W.-H. Chen, J.-S. Tsay, Strain driven phase
  transition and mechanism for \ce{Fe}/\ce{Ir} (111) films, Sci. Rep. 11~(1)
  (2021) 1--9.
\newblock \href {https://doi.org/10.1038/s41598-021-01474-1}
  {\path{doi:10.1038/s41598-021-01474-1}}.

\bibitem{https://doi.org/10.1002/zamm.201800179}
H.~Knüpfer, F.~Otto, Nucleation barriers for the cubic-to-tetragonal phase
  transformation in the absence of self-accommodation, ZAMM 99~(2) (2019)
  e201800179.
\newblock \href {https://doi.org/10.1002/zamm.201800179}
  {\path{doi:10.1002/zamm.201800179}}.

\bibitem{ZHANG20094332}
Z.~Zhang, R.~D. James, S.~Müller, Energy barriers and hysteresis in
  martensitic phase transformations, Acta Mater. 57~(15) (2009) 4332--4352.
\newblock \href {https://doi.org/10.1016/j.actamat.2009.05.034}
  {\path{doi:10.1016/j.actamat.2009.05.034}}.

\bibitem{CHEN20132566}
X.~Chen, V.~Srivastava, V.~Dabade, R.~D. James, Study of the cofactor
  conditions: Conditions of supercompatibility between phases, J. Mech. Phys.
  Solids 61~(12) (2013) 2566--2587.
\newblock \href {https://doi.org/10.1016/j.jmps.2013.08.004}
  {\path{doi:10.1016/j.jmps.2013.08.004}}.

\bibitem{Cui2006}
J.~Cui, Y.~S. Chu, O.~O. Famodu, Y.~Furuya, J.~Hattrick-Simpers, R.~D. James,
  A.~Ludwig, S.~Thienhaus, M.~Wuttig, Z.~Zhang, I.~Takeuchi, Combinatorial
  search of thermoelastic shape-memory alloys with extremely small hysteresis
  width, Nat. Mater. 5~(4) (2006) 286--290.
\newblock \href {https://doi.org/10.1038/nmat1593}
  {\path{doi:10.1038/nmat1593}}.

\bibitem{knupfer2013nucleation}
H.~Kn{\"u}pfer, R.~V. Kohn, F.~Otto, Nucleation barriers for the
  cubic-to-tetragonal phase transformation, Commun. pure appl. math. 66~(6)
  (2013) 867--904.

\bibitem{zhang2009energy}
Z.~Zhang, R.~D. James, S.~M{\"u}ller, Energy barriers and hysteresis in
  martensitic phase transformations, Acta Mater. 57~(15) (2009) 4332--4352.

\bibitem{balakrishna2022compatible}
A.~R. Balakrishna, Compatible microstructures in magnetic materials, Phys. Rev.
  Mater. 6~(7) (2022) 074402.
\newblock \href {https://doi.org/10.1103/PhysRevMaterials.6.074402}
  {\path{doi:10.1103/PhysRevMaterials.6.074402}}.

\bibitem{zhang2005phase}
J.~X. Zhang, L.~Q. Chen, Phase-field microelasticity theory and micromagnetic
  simulations of domain structures in giant magnetostrictive materials, Acta
  Mater. 53~(9) (2005) 2845--2855.
\newblock \href {https://doi.org/10.1016/j.actamat.2005.03.002}
  {\path{doi:10.1016/j.actamat.2005.03.002}}.

\bibitem{li2016giant}
Y.~Li, D.~Zhao, J.~Liu, Giant and reversible room-temperature elastocaloric
  effect in a single-crystalline \ce{NiFeGa} magnetic shape memory alloy, Sci.
  Rep. 6~(1) (2016) 1--11.
\newblock \href {https://doi.org/10.1038/srep25500}
  {\path{doi:10.1038/srep25500}}.

\bibitem{10.1063/1.3268471}
K.~E. Knipling, M.~Daniil, M.~A. Willard, {Fe-based nanocrystalline soft
  magnetic alloys for high-temperature applications}, Appl. Phys. Lett. 95~(22)
  (2009) 222516.
\newblock \href {https://doi.org/10.1063/1.3268471}
  {\path{doi:10.1063/1.3268471}}.

\bibitem{10.1063/1.5007248}
B.~Dong, J.~Healy, S.~Lan, M.~Daniil, M.~A. Willard, {Computational alloy
  design of \ce{(Co_{1-x}Ni_x)_{88}Zr_7B_4Cu_1} nanocomposite soft magnets},
  AIP Adv. 8~(5) (2018) 056124.
\newblock \href {https://doi.org/10.1063/1.5007248}
  {\path{doi:10.1063/1.5007248}}.

\bibitem{albertini2011}
F.~Albertini, S.~Fabbrici, A.~Paoluzi, J.~Kamarad, Z.~Arnold, L.~Righi,
  M.~Solzi, G.~Porcari, C.~Pernechele, D.~Serrate, P.~Algarabel, Reverse
  magnetostructural transitions by \ce{Co} and \ce{In} doping \ce{NiMnGa}
  alloys: Structural, magnetic, and magnetoelastic properties, Mater. Sci.
  Forum 684 (2011) 151--163.
\newblock \href {https://doi.org/10.4028/www.scientific.net/MSF.684.151}
  {\path{doi:10.4028/www.scientific.net/MSF.684.151}}.

\bibitem{Meisenheimer2021}
P.~B. Meisenheimer, R.~A. Steinhardt, S.~H. Sung, L.~D. Williams, S.~Zhuang,
  M.~E. Nowakowski, S.~Novakov, M.~M. Torunbalci, B.~Prasad, C.~J. Zollner,
  Z.~Wang, N.~M. Dawley, J.~Schubert, A.~H. Hunter, S.~Manipatruni, D.~E.
  Nikonov, I.~A. Young, L.~Q. Chen, J.~Bokor, S.~A. Bhave, R.~Ramesh, J.-M. Hu,
  E.~Kioupakis, R.~Hovden, D.~G. Schlom, J.~T. Heron, Engineering new limits to
  magnetostriction through metastability in iron-gallium alloys, Nat. Commun.
  12~(1) (2021) 2757.
\newblock \href {https://doi.org/10.1038/s41467-021-22793-x}
  {\path{doi:10.1038/s41467-021-22793-x}}.

\bibitem{9186272}
C.~Velez, D.~K. Patel, S.~Kim, M.~Babaei, C.~R. Knick, G.~L. Smith,
  S.~Bergbreiter, Hierarchical integration of thin-film niti actuators using
  additive manufacturing for microrobotics, J. Microelectromech. Syst. 29~(5)
  (2020) 867--873.
\newblock \href {https://doi.org/10.1109/JMEMS.2020.3019064}
  {\path{doi:10.1109/JMEMS.2020.3019064}}.

\bibitem{Yang2023}
Y.~Yang, T.~D. Oyedeji, X.~Zhou, K.~Albe, B.~Xu, Tailoring magnetic hysteresis
  of additive manufactured \ce{Fe-Ni} permalloy via multiphysics-multiscale
  simulations of process-property relationships, npj Comput. Mater. 9~(1)
  (2023) 103.
\newblock \href {https://doi.org/10.1038/s41524-023-01058-9}
  {\path{doi:10.1038/s41524-023-01058-9}}.

\bibitem{Chumak2021}
O.~M. Chumak, A.~Pacewicz, A.~Lynnyk, B.~Salski, T.~Yamamoto, T.~Seki, J.~Z.
  Domagala, H.~Głowiński, K.~Takanashi, L.~T. Baczewski, H.~Szymczak,
  A.~Nabiałek, Magnetoelastic interactions and magnetic damping in
  \ce{Co_2Fe_{0.4}Mn_{0.6}Si} and \ce{Co_2FeGa_{0.5}Ge_{0.5}} heusler alloys
  thin films for spintronic applications, Sci. Rep. 11~(1) (2021) 7608.
\newblock \href {https://doi.org/10.1038/s41598-021-87205-y}
  {\path{doi:10.1038/s41598-021-87205-y}}.

\bibitem{doi:10.1098/rsta.1948.0007}
E.~C. Stoner, E.~P. Wohlfarth, A mechanism of magnetic hysteresis in
  heterogeneous alloys, Philos. Transact. A Math. Phys. Eng. Sci. 240~(826)
  (1948) 599--642.
\newblock \href {https://doi.org/10.1098/rsta.1948.0007}
  {\path{doi:10.1098/rsta.1948.0007}}.

\bibitem{Kurti1988}
N.~Kurti (Ed.), Selected Works of Louis N\'eel, 1st Edition, CRC Press, 1988.
\newblock \href {https://doi.org/10.1201/9780367810580}
  {\path{doi:10.1201/9780367810580}}.

\bibitem{renuka2024Rethinking}
A.~Renuka~Balakrishna, Rethinking hysteresis in magnetic materials, MRS Commun.
  (2024).
\newblock \href {https://doi.org/10.1557/s43579-024-00624-6}
  {\path{doi:10.1557/s43579-024-00624-6}}.

\bibitem{10.1063/1.369539}
J.~Garc{\i}a-Otero, M.~Porto, J.~Rivas, A.~Bunde, Influence of the cubic
  anisotropy constants on the hysteresis loops of single-domain particles: A
  monte carlo study, Journal of Applied Physics 85~(4) (1999) 2287--2292.
\newblock \href {https://doi.org/10.1063/1.369539}
  {\path{doi:10.1063/1.369539}}.

\bibitem{PhysRevB.66.174419}
H.~Kachkachi, M.~Dimian, Hysteretic properties of a magnetic particle with
  strong surface anisotropy, Phys. Rev. B 66 (2002) 174419.
\newblock \href {https://doi.org/10.1103/PhysRevB.66.174419}
  {\path{doi:10.1103/PhysRevB.66.174419}}.

\bibitem{HERZER200599}
G.~Herzer,
  \href{https://www.sciencedirect.com/science/article/pii/S0304885305003513}{Anisotropies
  in soft magnetic nanocrystalline alloys}, Journal of Magnetism and Magnetic
  Materials 294~(2) (2005) 99--106, nANO2004.
\newblock \href {https://doi.org/https://doi.org/10.1016/j.jmmm.2005.03.020}
  {\path{doi:https://doi.org/10.1016/j.jmmm.2005.03.020}}.
\newline\urlprefix\url{https://www.sciencedirect.com/science/article/pii/S0304885305003513}

\bibitem{Khachaturyan1983}
A.~G. Khachaturyan, Theory of Structural Transformations in Solids, Wiley, New
  York, NY, 1983.

\end{thebibliography}

\newpage
\section*{\textcolor{black}{Supplementary information}}

\subsection*{Notations in our micromagnetics model}
\renewcommand{\arraystretch}{1.5}

\begin{table}[!h]
\centering
\begin{tabular}{p{2cm} p{12cm}}
\hline 
Notation & Description \tabularnewline
\hline
%Material constants & \tabularnewline
$\mathcal{E}$ & An ellipsoid magnetic body occupying a region $\mathcal{E}$ \tabularnewline

$\mathbb{R}^3$ & Three-dimensional space \tabularnewline

$\mu_0$ & Vacuum permeability constant \tabularnewline

$m_s$ & Saturation magnetization in the ferromagnetic phase \tabularnewline

$\mathbf{m}$ & Nondimensionalized magnetization, $\mathbf{m} = \frac{\mathbf{M}}{m_s}$ \tabularnewline

$\mathbf{A}=\mathrm{A}\mathbf{I}$ & Gradient energy (or exchange energy) coefficient \tabularnewline

$\kappa_1$ & Magnetocrystalline anisotropy constant associated with rotation the magnetization away from its easy axes  \tabularnewline

$\lambda_{100}$, $\lambda_{111}$  & Magnetostriction constants represent the relative change in length of a magnetic material when exposed to an external field. The subscripts denote the strains along the $\langle 100\rangle$ and the $\langle 111\rangle$ crystallographic directions.\tabularnewline

$\mathbf{E}$ & Strain tensor \tabularnewline

$\mathbf{E}_0(\vb{m})$ & Spontaneous (or preferred) strain tensor. For a cubic crystal it is calculated as:
\[  
\vb{E}_0(\vb{m}) = \dfrac{3}{2}\mqty(\lambda_{100}\mathrm{(m_1^2-1/3)} & \mathrm{\lambda_{111}m_1m_2} & \mathrm{\lambda_{111}m_1m_3} \\ \mathrm{\lambda_{111}m_1m_2} & \lambda_{100}\mathrm{(m_2^2-1/3)} & \mathrm{\lambda_{111}m_2m_3} \\ \mathrm{\lambda_{111}m_1m_3} & \mathrm{\lambda_{111}m_2m_3} & \lambda_{100}\mathrm{(m_3^2-1/3)}) \]

\tabularnewline

$\mathbb{C}$ & Elastic stiffness constant \tabularnewline

$\sigma_\mathrm{R}$ & Residual stress \tabularnewline

$\zeta_{\mathbf{m}}$ & Magnetostatic potential: 
\[
\zeta_{\mathbf{m}}(\vb{x}) = -\frac{1}{4\pi}\int_{\mathcal{E}} \grad \dfrac{1}{|\vb{x-y}|}\cdot \vb{M(y)}\dd \vb{y}\]
\vspace{-5mm}
\tabularnewline 

$\mathbf{H}_{\mathrm{d}}$ & Demagnetization field, $\mathbf{H}_{\mathrm{d}} = -\grad \zeta_{\mathbf{m}}$ \tabularnewline

$\mathbf{H}_{\mathrm{ext}}$ & External field \tabularnewline

\hline
\end{tabular}
\caption{Notations and description of symbols used in the text} \label{Symbols}
\end{table}

\newpage

\subsection*{Material parameters}
Table \ref{tab:materialconstant} shows the magnetic material constants, namely the anisotropy constant $\kappa_1$, the magnetostriction constants $\lambda_{100}, \lambda_{111}$, and the saturation magnetization $m_s$, of FeNi alloys used in this calculation. The elastic stiffness constants corresponding to the cubic symmetry of Fe-Ni alloys are used with $c_{11} = 240$ $\mathrm{GPa}$, $c_{12} = 89$ $\mathrm{GPa}$, and $c_{44} = 76$ $\mathrm{GPa}$. 

\begin{table}[h]
\begin{centering}
\textcolor{black}{}%
\begin{tabular}{ccccc}
\hline 
\textcolor{black}{Ni content ($\%$)} & \textcolor{black}{$\kappa_{1}$ (kJ/m$^3$)} & \textcolor{black}{$\lambda_{100}$ ($\times 10^{-6}$)} & \textcolor{black}{$\lambda_{111}$ ($\times 10^{-6}$)} & $\mathrm{m_s} \mathrm{(\times 10^{6} A/m)}$\tabularnewline
\hline 

\textcolor{black}{50} & 0.958 & 10.0 & 30.9 & 1.25 \tabularnewline
\textcolor{black}{78.5} & $-0.161$ & 11.8 & 1.91 & 0.84 \tabularnewline
\hline 
\end{tabular}\\
\par\end{centering}

\caption{\label{tab:materialconstant}List of material constants for the FeNi alloy system \cite{bozorth1953permalloy}.}
\end{table}

\subsubsection*{Homogeneous and Heterogeneous Strains}\label{Strain tensor-supplement}
The magnetoelastic energy in Eq.~\ref{MicromagneticsEnergy} accounts for the non-spontaneous strains $\mathbf{E} \neq \mathbf{E}_0(\mathbf{m})$ stored in the ellipsoid. These strains, on the one hand, arise from lattice misfit at the domain walls and defect-ellipsoid interfaces, and on the other hand arise from lattice distortions due to applied or residual stresses. Following Khachaturyan's theory \cite{Khachaturyan1983}, we decompose the total strain tensor on the ellipsoid as a sum of a homogeneous or a spatially independent component $\bar{\mathbf{E}}$, and a heterogeneous or a spatially varying component $\tilde{\mathbf{E}}(\mathbf{x})$: $\mathbf{E}=\tilde{\mathbf{E}} +\bar{\mathbf{E}}$.

The heterogeneous strain $\tilde{\mathbf{E}}(\mathbf{x})$ satisfies the mechanical equilibrium condition $\nabla\cdot\sigma=0$ (Eq.~\ref{eq:mechanicaleq}) and 
has non-zero values in the vicinity of the defect. The homogeneous strain $\bar{\mathbf{E}}$ represents the macroscopic shape change of the ellipsoid body and is defined such that:
\begin{align}
\int_{\mathcal{E}}\tilde{\mathbf{E}}\mathrm{d}\mathbf{x}=0. \label{HomogeneousStrain}
\end{align}
Using these definitions of the homogeneous and heterogeneous strains, we next rewrite the micromagnetics energy in Eq.~\ref{MicromagneticsEnergy} as:
\begin{align}
{\Psi} & =\int_{\mathcal{E}}\Bigg\{\nabla\mathbf{m}\cdot\mathbf{A}\nabla\mathbf{m}+\kappa_{1}(\mathrm{m}_{1}^{2}\mathrm{m}_{2}^{2}+\mathrm{m}_{2}^{2}\mathrm{m}_{3}^{2}+\mathrm{m}_{3}^{2}\mathrm{m}_{1}^{2})+\frac{1}{2}\bar{\mathbf{E}}\cdot\mathbb{C}\bar{\mathbf{E}}+ \bar{\mathbf{E}}\cdot\mathbb{C}\tilde{\mathbf{E}}-\bar{\mathbf{E}}\cdot\mathbb{C}[\mathbf{E_{0}\mathrm{(\mathbf{m})}}]\nonumber\\
& + \frac{1}{2}[\tilde{\mathbf{E}}-\mathbf{E}_{0}(\mathbf{m})]\cdot\mathbb{C}[\tilde{\mathbf{E}}-\mathbf{E}_{0}(\mathbf{m})]-\mu_{0}m_s\mathbf{H_{\mathrm{ext}}\cdot m} -\sigma_{\mathrm{R}}\cdot(\bar{\mathbf{E}}+\tilde{\mathbf{E}})\Bigg\}\mathrm{d\mathbf{x}} +\int_{\mathbb{R}^{3}}\frac{\mu_{0}}{2}\left|\mathbf{H}_\mathrm{d}\right|^{2}\mathrm{d\mathbf{x}}.\label{nondimensionalized}
\end{align}
To identify the homogeneous strain corresponding to the equilibrium state of the ellipsoid under an applied stress, we minimize the micromagnetics energy in Eq.~\ref{nondimensionalized} with respect to the homogeneous strain tensor $\bar{\mathbf{E}}$:
\begin{align}
\frac{\partial\Psi}{\partial\bar{\mathbf{E}}} & =\mathrm{V}\mathbb{C}\bar{\mathbf{E}}+\mathbb{C}\cancelto{0}{\int_{\mathcal{E}}\tilde{\mathbf{E}}\mathrm{d\mathbf{x}}}-\mathbb{C}\int_{\mathcal{E}}\mathbf{E}_{0}(\mathbf{m})\mathrm{d\mathbf{x}}-\mathrm{V}\sigma_{\mathrm{R}}=0.\label{stress-free-strain}
\end{align}
Here $\mathrm{V}$ is the volume of the ellipsoid. From Eq.~\ref{HomogeneousStrain}, we have $\int_{\mathcal{E}}\tilde{\mathbf{E}}\mathrm{d}\mathbf{x}=0$. Simplifying Eq.~\ref{stress-free-strain} the homogeneous strain on an ellipsoid, which is subjected to a finite and a homogeneous residual stress, is given by:
\begin{align}
\bar{\mathbf{E}} & =\frac{1}{\mathrm{V}}\int_{\mathcal{E}}\mathbf{E}_{0}(\mathbf{m})\mathrm{d\mathbf{x}}+\mathbb{C}^{-1}\sigma_{\mathrm{R}}\nonumber \\
& = \frac{1}{\mathrm{V}}\int_{\mathcal{E}}\mathbf{E}_{0}(\mathbf{m})\mathrm{d\mathbf{x}}+\mathbb{S}\sigma_{\mathrm{R}}\nonumber\\
& = \mathbf{E}_{0}(\bar{\mathbf{m}})+\mathbf{E}_\mathrm{R}.\label{homogeneous-residual-strain}
 \end{align}
In Eq.~\ref{homogeneous-residual-strain}, $\mathbb{S}$ represents the compliance tensor of the material, $\bar{\mathbf{m}}$ represents the volume average magnetization on the ellipsoid, and $\mathbf{E}_\mathrm{R}$ represents the homogeneous residual strain tensor corresponding to the homogeneous residual stresses $\sigma_{\mathrm{R}}$ applied to the ellipsoid $\mathcal{E}$. In our calculations (Fig.~\ref{ResidualStresses}), we apply uniaxial residual stress $\sigma_{\mathrm{R}} = \sigma_{11}\hat{\mathbf{e}}_1\otimes\hat{\mathbf{e}}_1$ of about 200 kPa. The corresponding strain response in the material can be approximated as 
\begin{equation}
    \mathbf{E}_\mathrm{R} = \varepsilon_\mathrm{R}\hat{\mathbf{e}}_1\otimes\hat{\mathbf{e}}_1 - \nu\varepsilon_\mathrm{R}\hat{\mathbf{e}}_2\otimes\hat{\mathbf{e}}_2 - \nu\varepsilon_\mathrm{R}\hat{\mathbf{e}}_3\otimes\hat{\mathbf{e}}_3, \label{eq:ER}
\end{equation}
where $\varepsilon_\mathrm{R} = S_{11}\sigma_{11} = \frac{c_{11}+c_{12}}{c_{11}^2+c_{11}c_{12}-2c_{12}^2}\sigma_{11}$ and $\nu = -S_{12}/S_{11} = c_{12}/(c_{11}+c_{12})$ is the Poisson's ratio. Since $c_{11},c_{12}\sim 10^2$ GPa and $\sigma_{11}\sim 10^2$ kPa, $\varepsilon_\mathrm{R}\sim 10^{-6} \ll 1$. These residual strains do not move the system significantly away from the bottom of the energy wells (i.e., the residual strains are negligible compared to the relatively large separation between the energy wells). These residual strains (or equivalently the residual stresses) affect the form of the parabolic locus as shown in Figs.~\ref{ResidualStresses}(a,c).
%However, we note that the actual value of $\varepsilon$ is also influenced by localized disturbances such as cavities or inclusion-type defects in the materials. 

\subsection*{Denominator of $\alpha$}
In Eq.~\ref{alpha_eval} we introduce a relation (denoted by $\alpha$) for which the energy barrier is minimized and is valid for non-zero values of its denominator, $\Delta F[\vb{m}] \ne 0$. In this subsection, we inspect the different values for $\Delta F[\vb{m}]$.

Let us start with a simplified case of an ellipsoid, in which the fields are uniform, and the magnetization $\vb{m} = \cos\theta\hat{\vb{e}}_1 + \sin\theta\hat{\vb{e}}_2$. Thus,
\begin{align}
    \Delta F[\vb{m}] &= V\Delta (\sin^2\theta\cos^2\theta) = V\dfrac{1}{2}\sin (4\theta) 
    \label{alpha-supplement}
\end{align}
In Eq.~\ref{alpha-supplement}, $\Delta F[\vb{m}] = 0$ would imply that $\theta = 0,~\pi/4,~\cdots$. However, for needle domains on an infinitely large ellipsoid, the perturbation to the average magnetization is $0<\theta\ll 1$. Therefore, without loss of generality, we assume that the denominator of $\alpha$ is typically greater than zero.

\subsection*{Parabolic Relation and Minimum Coercivity}\label{Shift-parabola}
In this subsection, we interpret the value of $k$ introduced in Eq.~\ref{parabolic-shift}. For micromagnetic calculations without residual stresses, e.g., in Fig.~\ref{CoercivityMaps}(a), the coercivity and the value of $k$ can be written as: 
\begin{equation}
\begin{aligned}
    &\Delta\Psi_{\mathrm{Zeeman}} = -\dfrac{\mu_0m_s|H_{\mathrm{c}}|}{\kappa_1}V\Delta (\cos\theta) = -\Delta\Psi_{\mathrm{tot}} \\
    \Rightarrow & |H_{\mathrm{c}}| = \dfrac{\kappa_1}{\mu_0m_sV}\dfrac{1}{\Delta (\cos\theta)}\Delta\qty[\int_{\mathcal{E}}\sum_{\substack{i,j=1 \\ i>j}}^3\mathrm{m}_i^2\mathrm{m}_j^2\dd\vb{x}+\frac{(c_{11}-c_{12})\lambda_{100}^2}{2\kappa_1}\int_{\mathcal{E}}\sum_{i=1}^3 \hat{\varepsilon}_{\mathrm{D},ii}^2\dd\vb{x}] \\
    \Rightarrow & k = \pdv[2]{|H_{\mathrm{c}}|}{\lambda_{100}} = \frac{c_{11}-c_{12}}{\mu_0m_sV}\dfrac{1}{\Delta (\cos\theta)}\Delta\qty(\int_{\mathcal{E}}\sum_{i=1}^3 \hat{\varepsilon}_{\mathrm{D},ii}^2\dd\vb{x}).
\end{aligned} \label{eval_k}
\end{equation}
Next, we consider a coercivity map with residual stresses of $\sigma_{11} = 200$ kPa as in Fig.~\ref{ResidualStresses}. On this coercivity map, we select three representative points at $\kappa_1 = 500, 1000, 1500~\mathrm{J/m^3}$ (with $c_{11}-c_{12} = 150$ GPa) with the corresponding magnetostriction constants $\lambda_{100}$ predicted by the parabolic locus of minimum coercivity Fig.~\ref{CoercivityMaps}(a): $\kappa_1 \approx \lambda_{100}^2/800$. We then calculate the values of $k$ and the parabolic shift $\beta\sigma_{11}$, respectively. We find that the estimated shift is close to what we observed in Fig.~\ref{ResidualStresses} ($\beta\sigma_{11} \approx 220\times 10^{-6}$) and it varies slowly with the changing $\lambda_{100}$, due to the minor changes in microstructures. We note that for large values of $\lambda_{100}$, such as 1500~$\mu\epsilon$, the shift is slightly larger and it deviates from the predicted minimum coercivity parabola.

\begin{table}[h]
\begin{centering}
\textcolor{black}{}%
\begin{tabular}{ccccc}
\hline 
\textcolor{black}{$\kappa_{1}$ (J/m$^3$)} & \textcolor{black}{$\lambda_{100}$ ($\times 10^{-6}$)} & \textcolor{black}{$k\mu_0m_s$ (GPa)} & $\beta\sigma_{11}$ ($\times 10^{-6}$)\tabularnewline
\hline 
 500 & 632.0 & 1.856 & 162 \tabularnewline 
1000 & 894.0 & 1.719 & 174 \tabularnewline
1500 & 1095.0 & 1.520 & 197 \tabularnewline
\hline 
\end{tabular}\\
\par\end{centering}

\caption{\label{tab:k-shift}Predicted value of $k$ and parabolic shift under residual stress $\sigma_{11} = 200$ kPa.}
\end{table}

Notice that our approximation $\vb{m}\approx (1,0,0)$ in the analysis of residual stress holds as long as the system is still in the needle domain state. In this state, the elastic energy decreases at magnetization reversal, which means $k>0$. Therefore, the minimum coercivity parabola is predicted to shift downward ($\sigma_{11} > 0$) or upward ($\sigma_{11} < 0$), and the trend of coercivities in Fig.~\ref{ResidualStresses} is the same as our prediction in Eq.~\ref{LinearShift}, exhibiting the anticipated increase (or decrease). In addition, we have observed that if $\sigma_{11}\lambda_{100}$ are the same, the coercivities are also the same in Fig.~\ref{ResidualStresses} (i.e. $H_{\mathrm{c}}(\kappa_1, \lambda_{100}, \sigma_{11}) = H_{\mathrm{c}}(\kappa_1, -\lambda_{100}, -\sigma_{11})$), as is also predicted by the symmetry of our free energy. On the other hand, if the residual stress is larger, the needle domain state might not be stable, and the coercivity maps could show quite distinct patterns.

\end{document}